\documentclass[final,5p,times,twocolumn]{elsarticle}
\biboptions{sort&compress}
\usepackage{lineno,hyperref}
\modulolinenumbers[5]
\usepackage{amsmath}
\usepackage{xcolor}
\usepackage{multirow}
\usepackage{listofsymbols}
\journal{ Test Journal of \LaTeX\ Templates}

\def\be{\begin{equation}}
\def\ee{\end{equation}}
\def\ba{\begin{eqnarray}}
\def\ea{\end{eqnarray}}

\begin{document}
\begin{frontmatter}
\title{A MULTI-FIELD TACHYON-QUINTOM MODEL OF DARK ENERGY AND FATE OF THE UNIVERSE}



\author[authorname]{Muhammad Zahid Mughal}
\ead{zahidmughal82@yahoo.com}
\author[authorname]{Iftikhar Ahmad}
\ead{dr.iftikhar@uog.edu.pk}
\begin{abstract}
We investigate a multi-field model of dark energy in this paper. We develop a model of dark energy with two multiple scalar fields, one we consider, is a multi-field tachyon and the other is multi-field phantom tachyon scalars. We make an analysis of the system in phase space by considering inverse square potentials suitable for these models. Through the development of an autonomous dynamical system, the critical points and their stability analysis is performed. It has been observed that these stable critical points are satisfied by power-law solutions. Moving on towards the analysis we can predict the fate of the universe. A special feature of this model is that it affects the equation of state parameter $w$  to alter from being it greater than $-1$  to be less than it during the evolutionary phase of the universe. Thus it’s all about the phantom divide which turns out to be decisive in the evolution of the cosmos in these models.
\end{abstract}
\end{frontmatter}

\section{Introduction}
The energy content of the universe and its nature is the basic issue and big challenge for today's cosmology both on theoretical and observational grounds. The accelerated expansion of the universe in the framework of general relativity necessitates the assumption of a hypothetical form of energy to be present in the universe. This energy density is usually known as dark energy and has negative pressure and is considered to accelerate the expansion of the universe. The vacuum energy with negative pressure has an interesting historical perspective which we discuss briefly here.
Einstein \cite{Pd1} was impelled to modify his field equations of general relativity to allow for a static model of the universe. For this purpose, he introduced a term known as cosmological constant denoted usually by the letter  $\text{ }\!\!\Lambda\!\!\text{ }$,  which characterizes the spatially homogeneous and flat, spherically symmetric, independent of time solution with positive constant curvature. $\text{ }\!\!\Lambda\!\!\text{ }$  represents constant repulsive-anti-gravity energy density ingrained into the structure of spacetime itself which counterbalances the gravity therewith entailing a static, non-dynamical universe \cite{Pd2}. In this context, Einstein was first to initiate Modern cosmology by applying the General Theory of Relativity to the whole geometry of spacetime of the universe and in rendering the use of vacuum energy as counteracting gravitational effects in the form of cosmological constant \cite{Pd3}. Shortly after Einstein’s static model of the universe, de Sitter presented a model of the universe that was supposed to present a static, flat universe without ordinary physical matter content contained in it. It was, later on, proved and observed that although de Sitter universe is empty of baryonic matter, despite this it expands which means that the geometry of spacetime is dynamic and does not necessarily require ordinary matter as Einstein believed following Mach's principle ( Weyl 1923, Lemaitre 1925). Therefore, only a positive cosmological constant $\text{ }\!\!\Lambda\!\!\text{ }$ drives the rate of expansion of the universe  $i.e.$  $H\propto \sqrt{\Lambda }$ \cite{Pd4, Pd5}.  $\Lambda$  has to be interpreted as representing dark energy or scalar field in later times. In the framework of general theory of relativity, the possible solutions for the dynamic universes were discussed and investigated earlier than any else by Alexander Friedmann (1888-1925) purely on theoretical grounds in \cite{Pd6, Pd7, Pd8}. These solutions do not require the use of the cosmological constant term $\Lambda$, although it was introduced by Einstein to represent the vacuum energy density inherent into the structure of spacetime itself, however, he never liked it as it mars the aesthetic beauty of the equations. Friedmann demonstrated that the field equations of general relativity admit non-static solutions and consequently the universe may contract, expand, collapse, and even be born from a singularity. He accordingly utilized three geometries of the universe namely spherical, hyperbolic, and flat. A Similar model indicating that our universe is expanding was developed by George Lemaitre in the coming years. Lemaitre was unaware of Friedmann’s work at that time and was motivated by observational evidence of the data collected by telescopes \cite{Pd9} afresh in those years. Basing his findings solely on observational data of redshifts of nebulae E.A. Hubble established the proof of expansion of the universe followed by showing the instability of Einstein’s static model of the universe by Eddington \cite{Pd10, Pd11}. Reluctantly convinced by contemporary developments of dynamical universe models Einstein abandoned the cosmological constant term calling it his blunder ever made in favor of dynamical universe and in the year 1931, he finally acceded to expanding universe \cite{Pd12}. Expanding universe had now been established and later developments about early universe emerged as the big bang model and later in the ’80s an exponential expansion period to be known as inflation was introduced by Starobinsky and Guth to remove the defects in the standard big bang cosmological model of the universe \cite{Pd13, Pd14}. In 1998 the work of S. Perlmutter, B. Schmidt, and A. Riess on accelerated expansion of the universe was propounded based on observational evidence, and it earned the 2011 Nobel prize for them \cite{Pd15, Pd16}. S. Perlmutter described the possibility of tracing out the history of the cosmological expansion of the universe and that the factor which is causing it to accelerate by considering the distant supernovae as standard candles. He also stated that accelerated expansion of the universe evidences the existence of dark energy but gives no clue of how can we identify it thereby establishing an acceleratingly expanding universe. Accelerated expansion of the universe posed a challenging problem to the model building of the universe. A large number of models have been constructed so far, for explaining the new form of the enigmatic energy which constitutes about 70\% of the universe but it still makes an open problem and is a vital area of research in contemporary theoretical and observational cosmologies.

In the recent past, based on the observational data collected it has been verified that the universe is in a state of accelerated expansion and proves to be spatially flat or very nearly to being spatially flat \cite{Pd17}. The energy-momentum tensor on the right-hand side of Einstein's field equations in the form of tensor representing the energy density of the normal matter in the universe succumbs to failure to take the responsibility of the talked accelerated expansion of the universe on the base of observational evidence. The status of  $\text{ }\!\!\Lambda\!\!\text{ }$  does not possess amenability and remains unacceptable due to fine-tuning problems associated with its value because it has a non-dynamical small constant value by terminological definition which suspects its candidacy for a true and feasible agent of the dark energy. The fields that are scalar in their nature can atone for the fine-tuning problems pertaining to  $\text{ }\!\!\Lambda\!\!\text{ }$  and might restitute to the insufficiency of the energy-momentum tensor as well. There is a large number of scalar field models like K-essence, quintessence, tachyon, phantom, phantom tachyon, etc. \cite{Pd18}. The quintessence scalar fields are armed with the properties of such behavior that can be likened to a tracker. Their nature of hinted at tracker-like modus operandi engenders it possible for the energy density contained in the fields to track the background energy densities before its prevailing dominance in the present scenario of the cosmos. In K-essence we do modification in the term of kinetical energy of the under-discussion acceleration which causes the expansion of the universe to accelerate. In the beginning, tachyon was a hypothetically considered particle propagating with a velocity faster than that of light. With the surfacing of string theory, its meaning was changed and it now represents varied states of fields quantum mechanically with imaginary masses \cite{Pd19}. General properties of the models based on the tachyon scalar field were studied in \cite{Pd20}. Although possessing instability, the tachyon field for certain potentials is believed to share its role vitally to be responsible for the late-time acceleration of the cosmos. A multifield tachyon model was considered to derive the acceleratedly expanding period of inflation by Y.S. Piao et al. \cite{Pd21}.

With the exception of phantom models of dark energy all the scalar field models represent an equation of state parameter  $\left( w \right)$  with value  $w\ge -1$. Rolling tachyon has an equation of state parameter whose value situates between  $-1$  and  $0$. There are multitudinous models based on rolling tachyon where an attempt has been made and it is inquired to construct an acceptable model incorporating inflation, dark energy, and dark matter. A condensate of a rolling tachyon is considered in string theories. The study of rolling tachyon has largely been carried out in \cite{Pd22} in connection with dark energy. The recent observational evidence on the base of data of the observable parameters suggest the value of the equation of state parameter  $w$  to be less than  $-1$  $i.e.$ $w<-1$  which is fulfilled by the phantom model of the dark energy. Phantom fields were historically introduced in the steady-state theory of the universe by Hoyle et al. Takin into account the transition to non-phantom theories in standard cosmology, phantom energy takes the responsibility of all radiation, matter, and structure formation in the universe and of its dynamics going back to the epochs of the very early universe where inflation occurs to the recently observed dark energy era. Big Rip singularity occurring in the future of the cosmos happens for the constant phantom condition. To evade this catastrophic Big Rip scenario some different models were proposed. Phantom dark energy with higher-order corrections to Einstein-Hilbert action with relation to the fixed dilaton and modulus fields was also studied in reference \cite{Pd23}.

The transition of the equation of state parameter  $~w$  across   $-1$  $i.e.$  terminologically known as phantom crossing or phantom divide is approved by an analytical study of the properties of the dark energy. This crossing of  $w$  across  $-1$  is satisfied by a scenario developed by quint-essence and phantom dark energy models known as quintom which is a newly coined word from the two contributing fields \cite{Pd24}.  The Quintom model of dark energy previously had an equation of state with $w>-1$ and recently it converted to  $w<-1$  which is in the agreement with current observational data. Quintom presents a dynamic model of dark energy. It differs from the cosmological constant, Quintessence, Phantom, K-essence, and so on in the determination of the cosmological evolution. A salient feature of the Quintom model is that its EoS can smoothly cross over w = -1.
In recent years there has been a lot of proposals for the Quintom-like models in the literature. Due to its quintom-like behavior, it was conveniently recognized and called as tachyon-quintom in \cite{Pd25}. The case of one tachyon and one phantom tachyon was investigated in the reference \cite{Pd26}. We consider the case of scalar multi fields and develop the model including multiple tachyons and multiple phantom tachyons. We shall observe that during the cosmological evolutionary development of the cosmos, the equation of state parameter  $w$  phantom-crosses and displays one of the most viable models of the dark energy falling close to the present-day observations.

 The design of the paper is laid out in the way as follows, section II is devoted to the development of the mathematics of the model. It does include the development of the autonomous dynamical system that plays a significant role in investigating and understanding the behavior of such models. We draw plots of evolving between the multi-scalar fields and parameters for the equation of state  $w$  and dark energy  ${{\Omega }_{DE}}$  as a function of the number of e-folds  $N$. In section III we make an analysis of the data and stability of the critical points inferring the future evolutionary development of the universe. Our discussion and conclusion are presented in section IV.

\section{Development of the mathematics of the model}
We take two homogeneous scalar fields namely multi-tachyon  $\mathop{\sum }\limits_{i=1}^{n} \xi _{i} $  and multi-phantom tachyon  $\mathop{\sum }\limits_{i=1}^{n} \eta _{i} $  which is also known as quintom \cite{Pd24}. Their corresponding potentials are $V\left(\mathop{\sum }\limits_{i=1}^{n} \xi _{i} \right)$ and  $V\left(\mathop{\sum }\limits_{i=1}^{n} \eta _{i} \right)$ respectively. In the background we consider the FLRW universe with four-dimensional flat spacetime. Since we shall use equation of state parameter $w$ expressed in terms of pressure and density as the ratio of the two, therefore we are going to take the fluid whose density is the function of pressure only $i.e.$ $\rho =\rho \left( p \right)$. The barotropic fluid is disseminated through and replenished in the universe with equation $p_{\gamma } =\left(\gamma -1\right)\; \rho _{\gamma } $ with the condition $0<\gamma \le 2,$ where $\gamma $ has different values for dust and radiation etc. The system of this type will have action of the form
\be S=\int \left(\frac{M_{Pl}^{2} R}{2} +\mathop{\sum }\limits_{i=1}^{n} {\rm {\mathcal L}}_{\xi _{i} } +\mathop{\sum }\limits_{j=1}^{n} {\rm {\mathcal L}}_{\eta _{j} } \; +{\rm {\mathcal L}}_{m} \right)\sqrt{-g} \; d^{4} x  \label{1a1}\ee

where

\be \mathop{\sum }\limits_{i=1}^{n} {\rm {\mathcal L}}_{\xi _{i} } =-V\left(\mathop{\sum }\limits_{i=1}^{n} \xi _{i} \right)\sqrt{1+g^{\mu \nu } \partial _{\mu } \left(\mathop{\sum }\limits_{i=1}^{n} \xi _{i} \right)\partial _{\nu } \left(\mathop{\sum }\limits_{i=1}^{n} \xi _{i} \right)}\label{1a2} \ee

and

\be \mathop{\sum }\limits_{j=1}^{n} {\rm {\mathcal L}}_{\eta _{i} } =-V\left(\mathop{\sum }\limits_{i=1}^{n} \eta _{i} \right)\sqrt{1-g^{\mu \nu } \partial _{\mu } \left(\mathop{\sum }\limits_{i=1}^{n} \eta _{i} \right)\partial _{\nu } \left(\mathop{\sum }\limits_{i=1}^{n} \eta _{i} \right)} \label{1a3} \ee
are the scalar field Langrangian densities where as  ${\rm {\mathcal L}}_{m} $ represents the Langrangian density of the matter fields as Langrangian.
The spatial homogeneity will imply  $\partial _{i} \mathop{\sum }\limits_{i=1}^{n} \xi _{i} =\partial _{i} \mathop{\sum }\limits_{i=1}^{n} \eta _{i} $, so that the solutions depend upon time only. The generalized energy densities of the fields result as
\be {\rho _{\left( {\mathop \sum \limits_{i = 1}^n {\xi _i}} \right)}} = \frac{{V\left( {\mathop \sum \nolimits_{i = 1}^n {\xi _i}} \right)}}{{\sqrt {1 - \mathop \sum \nolimits_{i = 1}^n {{\dot \xi }_i}^2} }}\label{1a4} \ee
\be {\rho _{\left( {\mathop \sum \limits_{i = 1}^n {\eta _i}} \right)}} = \frac{{V\left( {\mathop \sum \limits_{i = 1}^n {\eta _i}} \right)}}{{\sqrt {1 + \mathop \sum \limits_{i = 1}^n {{\dot \eta }_i}^2} }}\;\label{1a5} \ee
And the generalized pressures for both fields
\be \;{{p} _{\mathop {\mathop \sum \limits^n }\limits_{i = 1} {\mkern 1mu} {\xi _i}}} =  - V\left( {\mathop {\mathop \sum \limits^n }\limits_{i = 1} {\mkern 1mu} {\xi _i}} \right)\;\sqrt {1 - \mathop {\mathop \sum \limits^n }\limits_{i = 1} {\mkern 1mu} {{\dot \xi }_i}{{\mkern 1mu} ^2}} \label{1a6} \ee
\be \;{{p} _{\mathop {\mathop \sum \limits^n }\limits_{i = 1} {\mkern 1mu} {\eta _i}}} =  - V\left( {\mathop {\mathop \sum \limits^n }\limits_{i = 1} {\mkern 1mu} {\eta _i}} \right)\;\sqrt {1 + \mathop {\mathop \sum \limits^n }\limits_{i = 1} {\mkern 1mu} {{\dot \eta }_i}^2} \label{1a7} \ee
Now the equations of the scalar fields of generalized tachyon and generalized phantom tachyon read in following forms
\be \frac{{\sum\nolimits_{i = 1}^n {\ddot \xi } }}{{1 - {{\sum\nolimits_{i = 1}^n {{{\dot \xi }_i}} }^2}}} + 3H\sum\nolimits_{i = 1}^n {{{\dot \xi }_i}} {\mkern 1mu}  + \frac{{{V_{\sum\nolimits_{i = 1}^n {{\xi _i}} }}\left( {\sum\nolimits_{i = 1}^n {{\xi _i}} } \right)}}{{V\left( {\sum\nolimits_{i = 1}^n {{\xi _i}} } \right)}} = 0 \label{1a8} \ee
and
\be \frac{{\mathop \sum \limits_{i = 1}^n {{\ddot \eta }_i}}}{{1 - \mathop \sum \limits_{i = 1}^n {{\dot \eta }_i}^2}} + 3H\mathop {\mathop \sum \limits^n }\limits_{i = 1} {\mkern 1mu} {\dot \eta _i} - \frac{{{V_{\mathop \sum \limits_{i = 1}^n {\eta _i}}}\left( {\mathop \sum \limits_{i = 1}^n {\eta _i}} \right)}}{{V\left( {\mathop \sum \limits_{i = 1}^n {\eta _i}} \right)}} = 0 \label{1a9} \ee
And from equation of continuity, substituting $p_{\gamma } =\left(\gamma -1\right)\rho _{\gamma }$, we obtain
\be \frac{d\rho }{dt} +3H\left(\rho +p\right)=0 ,\label{1a10} \ee
\be \Rightarrow \; \; \; \; \; \; \; \frac{d\rho _{\gamma } }{dt} +3H\left(\rho _{\gamma } +p_{\gamma } \right)\label{1a11} \ee
\be \Rightarrow \; \; \; \; \; \; \; \; \frac{d\rho }{dt} -3H\gamma \rho _{\gamma } =0 \label{1a12} \ee
Now in order to find the equation for acceleration, we have to compute first the following two parameters $\dot{H}$ and $H^{2}$. We know from Friedmann's Equations that
\be H^{2} =\left(\frac{\dot{a}}{a} \right)^{2} =\frac{8\pi G}{3} \rho +\frac{k}{a^{2} } \label{1a13} \ee
\be \dot H + {H^2} = \frac{{\ddot a}}{a} =  - \frac{{4\pi G}}{3}\left( {\rho  + 3p} \right) \label{1a14} \ee
from above Eq. (\ref{1a13}) and Eq. (\ref{1a14}),  we have
\be \dot{H}=-\frac{4\pi G}{3} \left(\rho +p\right)\label{1a15} \ee
Now Eq. (\ref{1a15}) can be written for the densities and pressures of the generalized fields considered in the model and for the barotropic pressure and density, that is
\be \dot H =  - \frac{1}{{2M_{Pl}^2}}\left( {\;{\rho _{\mathop \sum \limits_{i = 1}^n {\xi _i}}} + \;{\rho _{\mathop \sum \limits_{i = 1}^n {\eta _i}}} + \;{p_{\mathop \sum \limits_{i = 1}^n {\xi _i}}} + \;{p_{\mathop \sum \limits_{i = 1}^n {\eta _i}}} + \;{\rho _\gamma } + \;{p_\gamma }} \right)\label{1a16} \ee
Now, substituting the values for the generalized energy densities and pressures, we have
\[\begin{array}{*{20}{l}}
{\dot H =  - \frac{1}{{2M_{Pl}^2}}\left[ {\left( {\frac{{V\left( {\sum\nolimits_{i = 1}^n {{\xi _i}} } \right)}}{{\sqrt {1 - \sum\nolimits_{i = 1}^n {{{\dot \xi }_i}^2} } }} + \frac{{V\left( {\left( {\sum\nolimits_{i = 1}^n {{\eta _i}} } \right)} \right)}}{{\sqrt {1 - \sum\nolimits_{i = 1}^n {{{\dot \eta }_i}^2} } }}} \right)} \right.}\\
{\left. { - \left( \begin{array}{l}
V\left( {\sum\limits_{i = 1}^n {{\xi _i}} } \right)\;\sqrt {1 - \sum\limits_{i = 1}^n {{{\dot \xi }_i}^2} }  - V\left( {\sum\limits_{i = 1}^n {{\eta _i}} } \right)\;\sqrt {1 - \sum\limits_{i = 1}^n {{{\dot \eta }^2}} } \\
 + \;{\rho _\gamma } + \left( {\gamma  - 1} \right){\rho _\gamma }
\end{array} \right)} \right]}
\end{array}\]
after having simplified the above equation, we obtain
\be \dot{H}=-\frac{1}{2M_{Pl}^{2} } \left(\frac{\mathop{\sum }\nolimits_{i}^{n} \dot{\xi }_{i} ^{2} V\left(\mathop{\sum }\nolimits_{i=1}^{n} \xi _{i} \right)}{\sqrt{1-\mathop{\sum }\nolimits_{i=1}^{n} \dot{\xi }_{i} ^{2} } } +\frac{\mathop{\sum }\nolimits_{i}^{n} \dot{\eta }_{i} ^{2} V\left(\mathop{\sum }\nolimits_{i=1}^{n} \eta _{i} \right)}{\sqrt{1-\mathop{\sum }\nolimits_{i=1}^{n} \dot{\eta }_{i} ^{2} } } +\gamma \rho _{\gamma } \right)\label{1a17} \ee
Now from Eq. (\ref{1a13})
\[H^{2} =\left(\frac{\dot{a}}{a} \right)^{2} =\frac{8\pi G}{3} \rho +\frac{k}{a^{2} } \]
With $k=0$ for flat universe
\be H^{2} =\frac{8\pi G}{3} \rho \label{1a18} \ee
since $\rho =\rho \left( \underset{i=1}{\overset{n}{\mathop{\sum }}}\,{{\xi }_{i}},\ \underset{i=1}{\overset{n}{\mathop{\sum }}}\,{{\eta }_{i}},\gamma  \right)$, therefore
\be H^{2} =\frac{8\pi G}{3} \rho \left(\mathop{\sum }\limits_{i=1}^{n} \xi _{i} ,\; \mathop{\sum }\limits_{i=1}^{n} \eta _{i} ,\gamma \right)\label{1a19} \ee
\be H^{2} =\frac{1}{3M_{Pl}^{2} } \rho \left(\xi ,\; \eta ,\gamma \right)=\frac{1}{3M_{Pl}^{2} } \left(\; \rho _{\mathop{\sum }\limits_{i=1}^{n} \xi _{i} } +\; \rho _{\mathop{\sum }\limits_{i=1}^{n} \eta _{i} } +\rho _{\gamma } \right)\label{1a20} \ee
\be H^{2} =\frac{1}{3M_{Pl}^{2} } \left(\frac{V\left(\left(\mathop{\sum }\nolimits_{i=1}^{n} \xi _{i} \right)\right)}{\sqrt{1-\mathop{\sum }\nolimits_{i=1}^{n} \dot{\xi }_{i} ^{2} } } +\frac{V\left(\mathop{\sum }\nolimits_{i=1}^{n} \eta _{i} \right)}{\sqrt{1+\mathop{\sum }\nolimits_{i=1}^{n} \dot{\eta }_{i} ^{2} } } +\rho _{\gamma } \right)\label{1a21} \ee
Now dividing the both sides by $H^{2}$
\be 1=\frac{1}{3H^{2} M_{Pl}^{2} } \left(\frac{V\left(\mathop{\sum }\nolimits_{i=1}^{n} \xi _{i} \right)}{\sqrt{1-\mathop{\sum }\nolimits_{i=1}^{n} \dot{\xi }_{i} ^{2} } } +\frac{V\left(\mathop{\sum }\nolimits_{i=1}^{n} \eta _{i} \right)}{\sqrt{1+\mathop{\sum }\nolimits_{i=1}^{n} \dot{\eta }_{i} ^{2} } } +\rho _{\gamma } \right)\label{1a22} \ee
\be 1=\left(\frac{V\left(\mathop{\sum }\nolimits_{i=1}^{n} \xi _{i} \right)\backslash 3H^{2} M_{Pl}^{2} }{\sqrt{1-\mathop{\sum }\nolimits_{i=1}^{n} \dot{\xi }_{i} ^{2} } } +\frac{V\left(\mathop{\sum }\nolimits_{i=1}^{n} \eta _{i} \right)\backslash 3H^{2} M_{Pl}^{2} }{\sqrt{1+\mathop{\sum }\nolimits_{i=1}^{n} \dot{\eta }_{i} ^{2} } } +\frac{\rho _{\gamma } }{3H^{2} M_{Pl}^{2} } \right)\label{1a23} \ee
 From Eq. (\ref{1a23}), we are going to define some parameters which are dimensionless and can facilitate our computations
\be \; x_{\mathop{\sum }\limits_{i=1}^{n} \xi _{i} } =\mathop{\sum }\limits_{i=1}^{n} \dot{\xi }_{i} \label{1a24} \ee
\be \; x_{\mathop{\sum }\limits_{i=1}^{n} \eta _{i} } =\mathop{\sum }\limits_{i=1}^{n} \dot{\eta }_{i} \label{1a25} \ee
and
\be \; y_{\mathop{\sum }\limits_{i=1}^{n} \xi _{i} } =\frac{V\left(\mathop{\sum }\nolimits_{i=1}^{n} \xi _{i} \right)}{3H^{2} M_{Pl}^{2} } \label{1a26} \ee
\be \; y_{\mathop{\sum }\limits_{i=1}^{n} \eta _{i} } =\frac{V\left(\mathop{\sum }\nolimits_{i=1}^{n} \eta _{i} \right)}{3H^{2} M_{Pl}^{2} } \label{1a27} \ee
and
\be z=\frac{\rho _{\gamma } }{3H^{2} M_{Pl}^{2}} \label{1a28} \ee
The acceleration equation can be obtained from Eq. (\ref{1a17}) and Eq. (\ref{1a21}).
\be \begin{array}{l}
\frac{{\dot H}}{{{H^2}}} = \frac{{HH'}}{{{H^2}}} = \frac{{H'}}{H}\\
 =  - \frac{3}{2}\left( { - \frac{{\;{y_{\left( {\sum _{i = 1}^n{\xi _i}} \right)}}(\gamma  - x\sum _{i = 1}^n\xi _i^2}}{{\sqrt {1 - \sum _{i = 1}^n\dot \xi _i^2} }} - \frac{{\;{y_{\left( {\sum _{i = 1}^n{\eta _i}} \right)}}(\gamma  - x\sum _{i = 1}^n\eta _i^2}}{{\sqrt {1 + \sum _{i = 1}^n\dot \eta _i^2} }} + \gamma } \right)
\end{array}\label{1a29} \ee
Here $\mathop{H}\limits^{{'} } $ denotes the derivative of $H$ with respect to the logarithm of the scale factor $i.e.$ $lna=N.$
Now the Eq. (\ref{1a23}) becomes
\be 1=\left(\frac{\; y_{\mathop{\sum }\nolimits_{i=1}^{n} \xi _{i} } }{\sqrt{1-\; x_{\mathop{\sum }\nolimits_{i=1}^{n} \xi _{i} } ^{2} } } +\frac{\; y_{\mathop{\sum }\nolimits_{i=1}^{n} \eta _{i} } }{\sqrt{1+\; x_{\mathop{\sum }\nolimits_{i=1}^{n} \eta _{i} } ^{2} } } +z\right)\label{1a30} \ee
\be \Rightarrow 1=\left({\it }_{DE} +z\right)\label{1a31} \ee
where
\be {\it }_{DE} =\frac{\; y_{\mathop{\sum }\nolimits_{i=1}^{n} \xi _{i} } }{\sqrt{1-\; x_{\mathop{\sum }\nolimits_{i=1}^{n} \xi _{i} } ^{2} } } +\frac{\; y_{\mathop{\sum }\nolimits_{i=1}^{n} \eta _{i} } }{\sqrt{1+\; x_{\mathop{\sum }\nolimits_{i=1}^{n} \eta _{i} } ^{2} } } \label{1a32} \ee
The parameter ${\Omega}_{DE} $ weighs out the dark energy as a fraction of the critical density ${\Omega}_{CR}$
The equation of state parameter $w$ for dark energy of the system of scalars is given by
\be w=\frac{\rho _{\mathop{\sum }\nolimits_{i=1}^{n} \xi _{i} } +\rho _{\mathop{\sum }\nolimits_{i=1}^{n} \eta _{i} } }{p_{\mathop{\sum }\nolimits_{i=1}^{n} \xi _{i} } +p_{\mathop{\sum }\nolimits_{i=1}^{n} \eta _{i} } } \label{1a33} \ee

\be w=\frac{-V\left(\mathop{\sum }\nolimits_{i}^{n} \xi _{i} \right)\sqrt{1-\mathop{\sum }\nolimits_{i=1}^{n} \dot{\xi }_{i} ^{2} } -V\left(\mathop{\sum }\nolimits_{i=1}^{n} \eta _{i} \right)\; \sqrt{1+\mathop{\sum }\nolimits_{i=1}^{n} \dot{\eta }_{i} ^{2} } }{\frac{V\left(\mathop{\sum }\nolimits_{i=1}^{n} \xi _{i} \right)}{\sqrt{1-\mathop{\sum }\nolimits_{i=1}^{n} \dot{\xi }_{i} ^{2} } } +\frac{V\left(\mathop{\sum }\nolimits_{i=1}^{n} \eta _{i} \right)}{\sqrt{1+\mathop{\sum }\nolimits_{i=1}^{n} \dot{\eta }_{i} ^{2} } } } \label{1a34} \ee
dividing and multiplying by $\frac{8\pi G}{3H^{2} }$
\be w=\frac{\frac{8\pi G}{3H^{2} } \left(-V\left(\mathop{\sum }\nolimits_{i=1}^{n} \xi _{i} \right)\sqrt{1-\mathop{\sum }\nolimits_{i=1}^{n} \dot{\xi }_{i} ^{2} } -V\left(\mathop{\sum }\nolimits_{i=1}^{n} \eta _{i} \right)\sqrt{1-\mathop{\sum }\nolimits_{i=1}^{n} \dot{\eta }_{i} ^{2} } \right)}{\frac{8\pi G}{3H^{2} } \left(\frac{V\left(\mathop{\sum }\nolimits_{i=1}^{n} \xi _{i} \right)}{\sqrt{1-\mathop{\sum }\nolimits_{i=1}^{n} \dot{\xi }_{i} ^{2} } } +\frac{V\left(\mathop{\sum }\nolimits_{i=1}^{n} \eta _{i} \right)}{\sqrt{1-\mathop{\sum }\nolimits_{i=1}^{n} \dot{\eta }_{i} ^{2} } } \right)} \label{1a35} \ee
after simplification we get
\be w=\frac{-\; y_{\mathop{\sum }\nolimits_{i=1}^{n} \xi _{i} } \sqrt{1-\; x_{\mathop{\sum }\nolimits_{i=1}^{n} \xi _{i} } ^{2} } -\; y_{\mathop{\sum }\nolimits_{i=1}^{n} \eta _{i} } \sqrt{1-\; x_{\mathop{\sum }\nolimits_{i=1}^{n} \eta _{i} } ^{2} } }{\frac{y_{\mathop{\sum }\nolimits_{i=1}^{n} \xi _{i} } }{\sqrt{1-\; x_{\mathop{\sum }\nolimits_{i=1}^{n} \xi _{i} } ^{2} } } +\frac{y_{\mathop{\sum }\nolimits_{i=1}^{n} \eta _{i} } }{\sqrt{1-\; x_{\mathop{\sum }\nolimits_{i=1}^{n} \eta _{i} } ^{2} } } } \label{1a36} \ee

\begin{figure*}[ht!]
\centering
\begin{center}
  \includegraphics[scale=0.50]{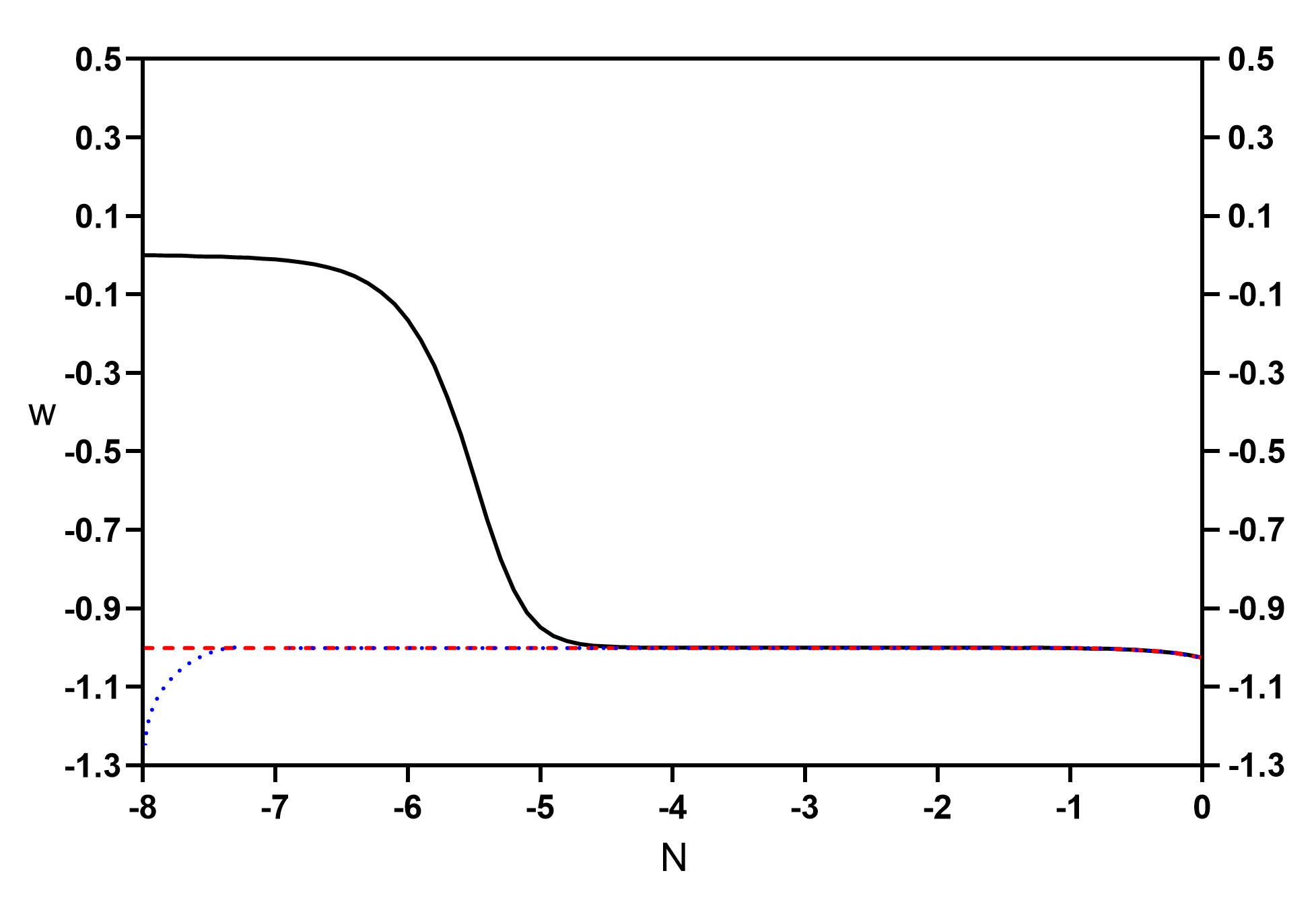}
  \includegraphics[scale=0.50]{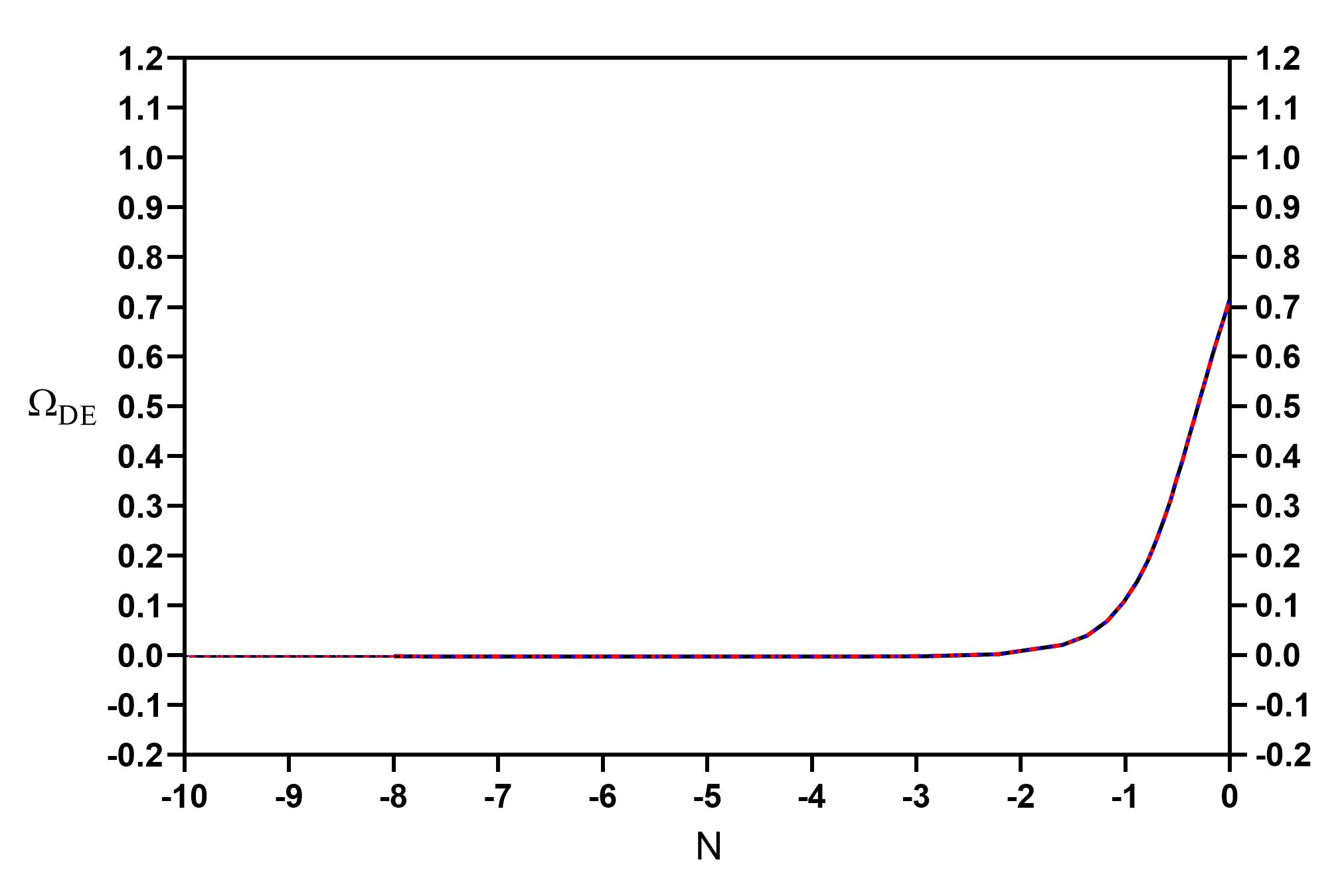}
\caption{The figure shows how does the growth of equation of state parameter $w$ and the parameter of dark energy density ${{\Omega }_{DE}}$ occurs with evolution of the number of e-folds $N$ and for $\gamma =1$ and ${{\beta }_{\sum\limits_{i=1}^{n}{{{\xi }_{i}}}}}$ and ${{\beta }_{\sum\limits_{i=1}^{n}{{{\eta }_{i}}}}}$ as 0.33.} \label{Figure: Fig1}
\end{center}
\end{figure*}

Now we develop an autonomous dynamical system from the Eq. (\ref{1a8}) and  Eq. (\ref{1a9}) for evolution of the system
\be\begin{array}{l}
{{x'}_{\mathop \sum \limits_{i = 1}^n {\xi _i}}} = \frac{d}{{dN}}\left( {{x_{\mathop \sum \limits_{i = 1}^n {\xi _i}}}} \right)\\
 =  - 3\left( {1 - \;x_{\mathop \sum \limits_{i = 1}^n {\xi _i}}^2} \right)\left( {{x_{\mathop \sum \limits_{i = 1}^n {\xi _i}}} - \sqrt {{\lambda _{\mathop \sum \limits_{i = 1}^n {\xi _i}}}{y_{\mathop \sum \limits_{i = 1}^n {\xi _i}}}} } \right)
\end{array}\label{1a37} \ee
\be \begin{array}{l}
{{y'}_{\mathop \sum \limits_{i = 1}^n {\xi _i}}} = \frac{d}{{dN}}\left( {{y_{\mathop \sum \limits_{i = 1}^n {\xi _i}}}} \right)\\
 = 3\;{y_{\sum\limits_{i = 1}^n {{\xi _i}} }}\left( {\begin{array}{*{20}{l}}
{ - \frac{{\;{y_{\sum\nolimits_{i = 1}^n {{\xi _i}} }}\left( {\gamma  - \;{x_{\sum\nolimits_{i = 1}^n {{\xi _i}} }}^2} \right)}}{{\sqrt {1 - \;{x_{\sum\nolimits_{i = 1}^n {{\xi _i}} }}^2} }} - \frac{{\;{y_{\sum\nolimits_{i = 1}^n {{\eta _i}} }}\left( {\gamma  - \;{x_{\sum\nolimits_{i = 1}^n {{\eta _i}} }}^2} \right)}}{{\sqrt {1 + \;{x_{\sum\nolimits_{i = 1}^n {{\eta _i}} }}^2} }}}\\
{ - \sqrt {{\lambda _{\sum\limits_{i = 1}^n {{\xi _i}} }}{y_{\sum\limits_{i = 1}^n {{\xi _i}} }}} {x_{\sum\limits_{i = 1}^n {{\xi _i}} }} + \gamma }
\end{array}} \right)
\end{array}\label{1a38} \ee
where
\be \lambda _{\mathop{\sum }\limits_{i=1}^{n} \xi _{i} } =\frac{4M_{PL}^{2} }{3M_{\mathop{\sum }\nolimits_{i=1}^{n} \xi _{i} }^{2} } \label{1a39} \ee
Now,
\be x'_{\mathop{\sum }\limits_{i=1}^{n} \eta _{i} } =\frac{d}{dN} x_{\mathop{\sum }\limits_{i=1}^{n} \eta _{i} } =-3\left(1+\; x_{\mathop{\sum }\limits_{i=1}^{n} \eta _{i} } ^{2} \right)\left(x_{\mathop{\sum }\limits_{i=1}^{n} \eta _{i} } +\sqrt{\lambda _{\mathop{\sum }\limits_{i=1}^{n} \eta _{i} } y_{\mathop{\sum }\limits_{i=1}^{n} \eta _{i} } } \right)\label{1a40}\ee
and
 $y'_{\mathop{\sum }\limits_{i=1}^{n} \eta _{i} } =\frac{d}{dN} y_{\mathop{\sum }\limits_{i=1}^{n} \eta _{i} }$
\be \begin{array}{r}
 = 3\;{y_{\mathop \sum \limits_{i = 1}^n {\eta _i}}}\left[ { - \frac{{\;{y_{\sum _{i = 1}^n{\xi _i}}}\left( {\gamma  - \;x_{\sum _{i = 1}^n{\xi _i}}^2} \right)}}{{\sqrt {1 - \;x_{\sum _{i = 1}^n{\xi _i}}^2} }} - \frac{{\;{y_{\sum _{i = 1}^n{\eta _i}}}\left( {\gamma  + \;x_{\sum _{i = 1}^n{\eta _i}}^2} \right)}}{{\sqrt {1 + \;x_{\sum _{i = 1}^n{\eta _i}}^2} }}} \right.\\
\left. { - \sqrt {{\lambda _{\mathop \sum \limits_{i = 1}^n {\eta _i}}}{y_{\mathop \sum \limits_{i = 1}^n {\eta _i}}}} {x_{\mathop \sum \limits_{i = 1}^n {\eta _i}}} + \gamma } \right]
\end{array}\label{1a41}\ee
where
\be \lambda _{\mathop{\sum }\limits_{i=1}^{n} \eta _{i} } =\frac{4M_{PL}^{2} }{3M_{\mathop{\sum }\nolimits_{i=1}^{n} \eta _{i} }^{2} } \label{1a42}\ee
\begin{figure}[htp!]
\centering
\begin{center}
  \includegraphics[scale=0.48]{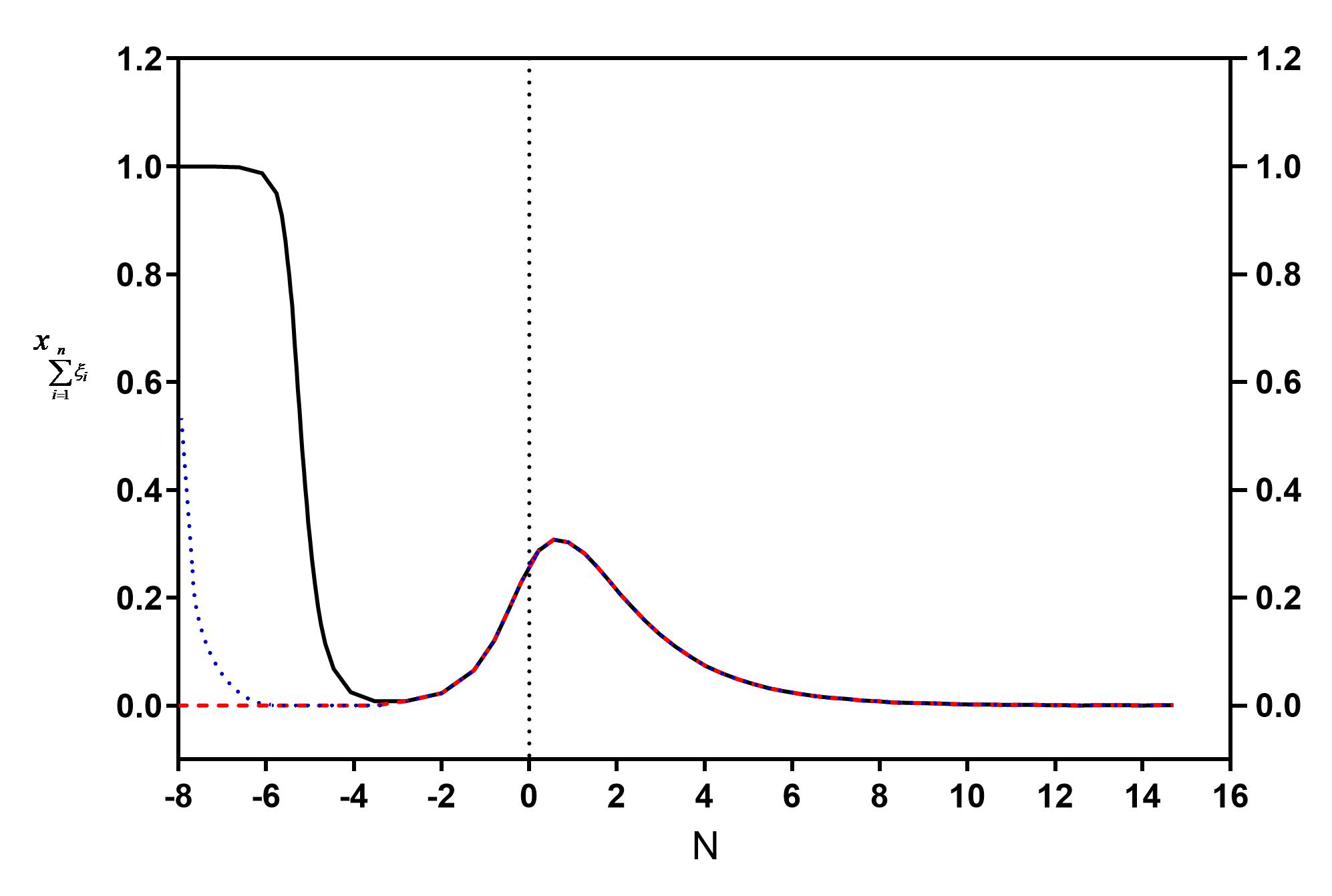}
  \includegraphics[scale=0.48]{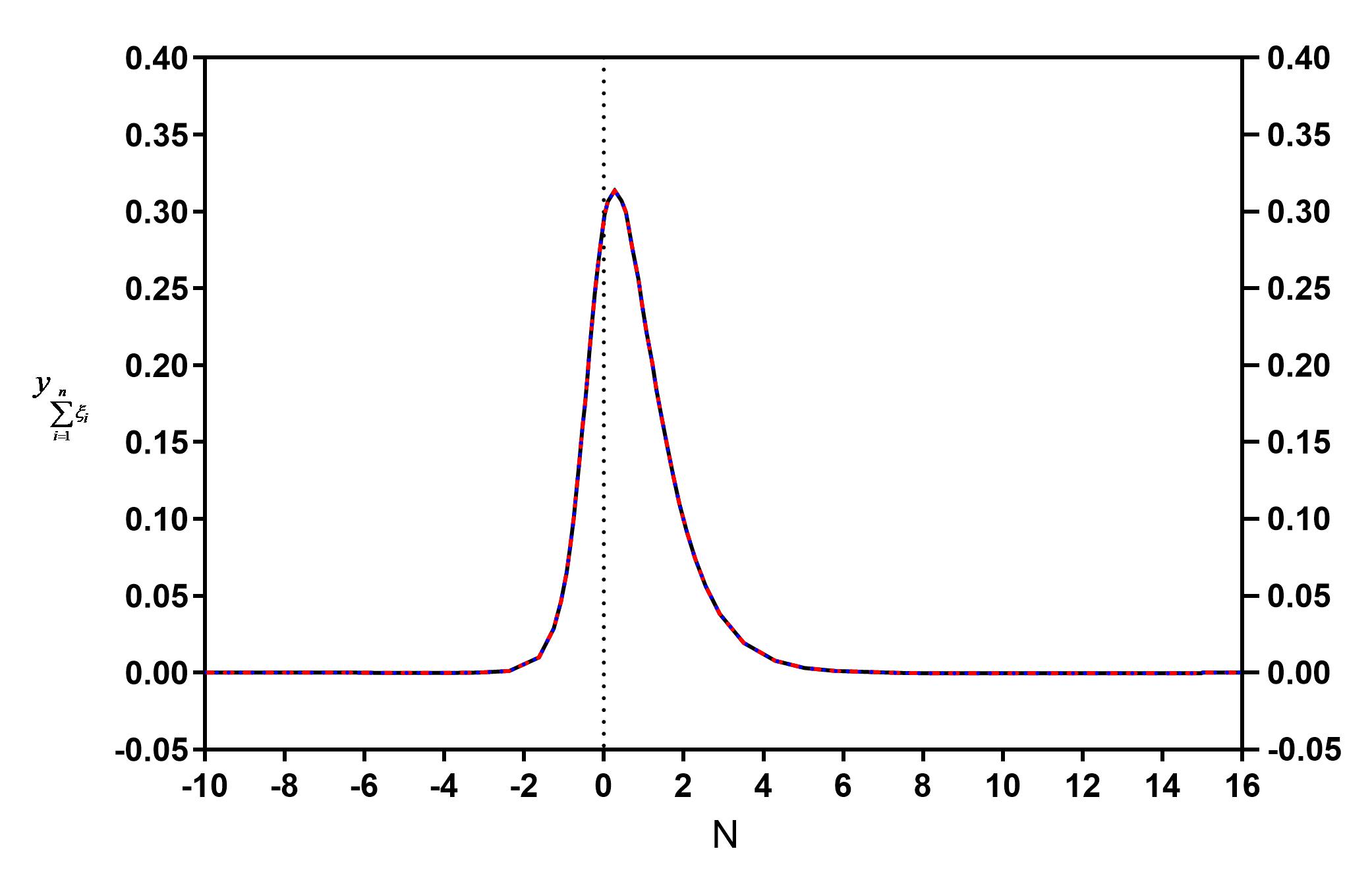}
  \includegraphics[scale=0.48]{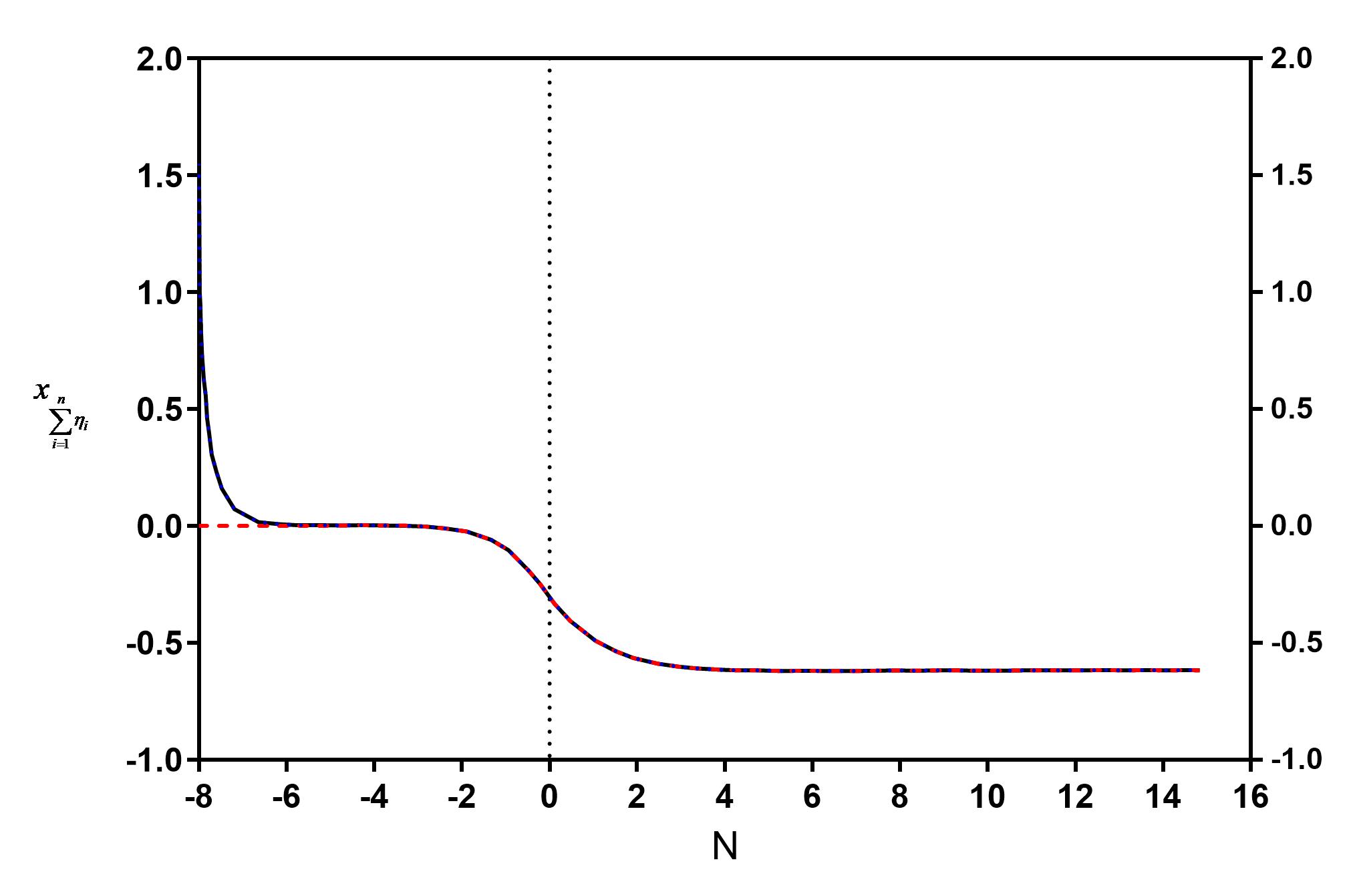}
  \includegraphics[scale=0.48]{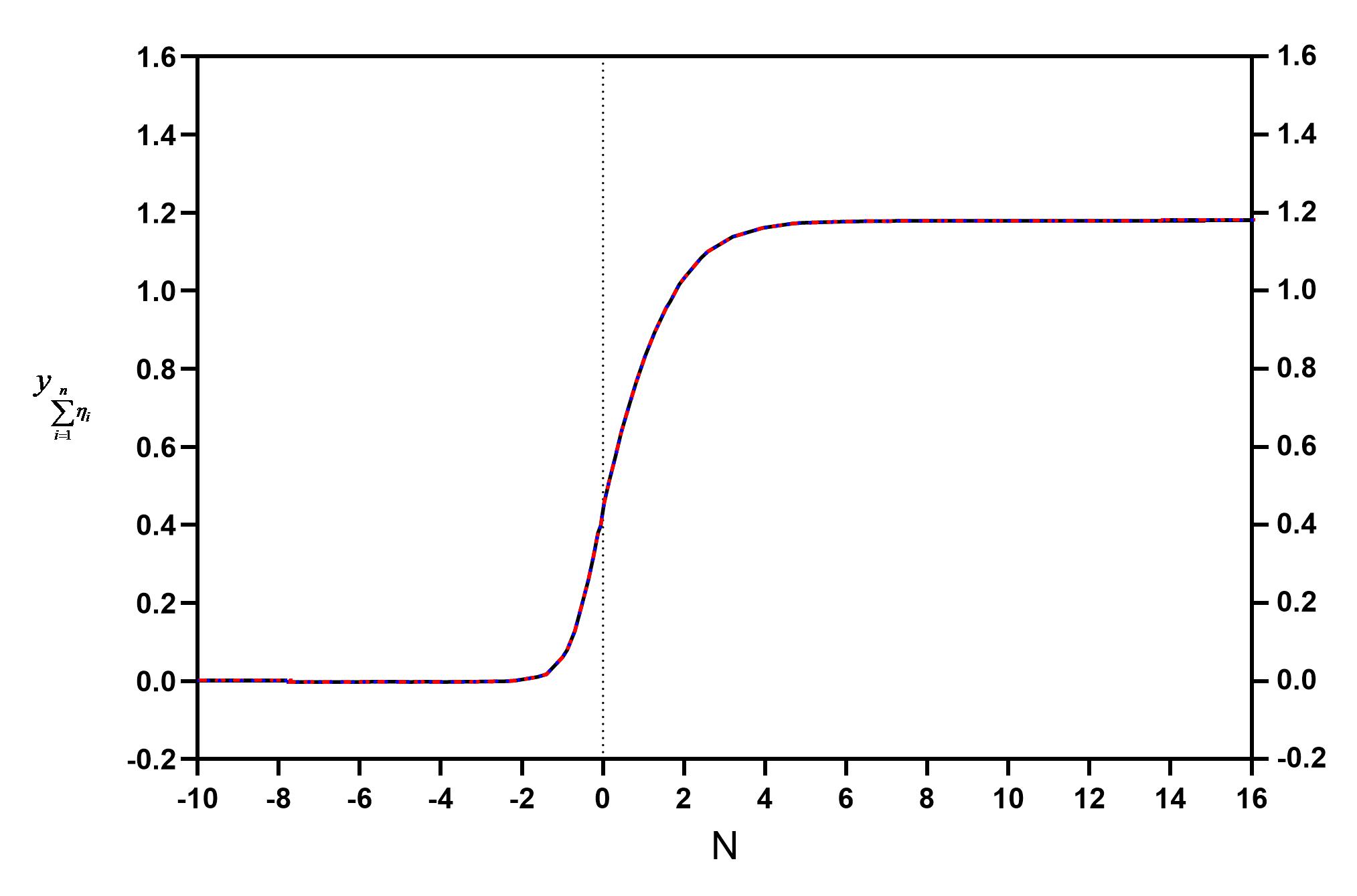}
\caption{The above figures indicate the behavior of general points of scalar multifields as the number of e-folds $N$ evolve. The points $x_{\mathop{\sum }\limits_{i=1}^{n} \xi _{i} } ,\; \; y_{\mathop{\sum }\limits_{i=1}^{n} \xi _{i} } ,\; \; x_{\mathop{\sum }\limits_{i=1}^{n} \eta _{i} } ,\; $and  $\; y_{\mathop{\sum }\limits_{i=1}^{n} \eta _{i} }$ develop gradually as the function of e-folding number $N$ for $\gamma =1$ and ${{\beta } {\sum\limits_{i=1}^{n}{{{\xi }_{i}}}}}$ and ${{\beta }_{\sum\limits_{i=1}^{n}{{{\eta }_{i}}}}}$ both with assigned a value equivalent to 0.33.} \label{Figure: Fig2}
\end{center}
\end{figure}

\section{Stability of the model}
Where the fixed points  $x'_{\mathop{\sum }\limits_{i=1}^{n} \xi _{i} } ,\; \; y'_{\mathop{\sum }\limits_{i=1}^{n} \xi _{i} } ,\; \; x'_{\mathop{\sum }\limits_{i=1}^{n} \eta _{i} } \; $and $y'_{\mathop{\sum }\limits_{i=1}^{n} \eta _{i} } $ diminish to zero, the existence of these critical points  ${\it x}_{\left(\mathop{\sum }\limits_{{\it i}=1}^{{\it n}} {\it \xi }_{{\it i}} \right)crt} $, ${\it y}_{\left(\mathop{\sum }\limits_{{\it i}=1}^{{\it n}} {\it \xi }_{{\it i}} \right)crt} $, ${\it x}_{\left(\mathop{\sum }\limits_{{\it i}=1}^{{\it n}} {\it \eta }_{{\it i}} \right)crt} $  and ${\it \; y}_{\left(\mathop{\sum }\limits_{{\it i}=1}^{{\it n}} {\it \eta }_{{\it i}} \right)crt}$ corresponds there. These critical points have been calculated and are listed in the Table \ref{Tab:tab1} below.
\begin{table*}[ht!]
\centering
\caption{enlisting the critical points}\label{Tab:tab1}
\begin{tabular}{cccccc}
\hline
\hline
\\
Sr.no. & ${\it x}_{\left(\mathop{\sum }\limits_{{\it i}=1}^{{\it n}} {\it \xi }_{{\it i}} \right)crt} $ & ${\it y}_{\left(\mathop{\sum }\limits_{{\it i}=1}^{{\it n}} {\it \xi }_{{\it i}} \right)crt} $ & ${\it x}_{\left(\mathop{\sum }\limits_{{\it i}=1}^{{\it n}} {\it \eta }_{{\it i}} \right)crt} $ & ${\it \; y}_{\left(\mathop{\sum }\limits_{{\it i}=1}^{{\it n}} {\it \eta }_{{\it i}} \right)crt} $ & Existence Status \\ \\
\hline
I & 0 & 0 & 0 & 0 & $\mathop{\sum }\limits_{j>0}^{N=2} \gamma _{j} $ \\
II & 0 & 0 & $-\sqrt{{\it \lambda }_{\mathop{\sum }\limits_{{\it i}=1}^{{\it n}} {\it \eta }_{{\it i}} } {\it y}_{\left(\mathop{\sum }\limits_{{\it i}=1}^{{\it n}} {\it \eta }_{{\it i}} \right)crt} } $ & $\frac{\sqrt{\lambda _{\left(\mathop{\sum }\nolimits_{{\it i}=1}^{{\it n}} {\it \eta }_{{\it i}} \right)} ^{2} +4} -\lambda _{\left(\mathop{\sum }\nolimits_{{\it i}=1}^{{\it n}} {\it \xi }_{{\it i}} \right)} }{2} $ & $\mathop{\sum }\limits_{j>0}^{N=2} \gamma _{j} $ \\
III & $\pm 1$ & 0 & 0 & 0 & $\mathop{\sum }\limits_{j>0}^{N=2} \gamma _{j} $ \\
IV & $\pm 1$ & 0 & $-\sqrt{{\it \lambda }_{\mathop{\sum }\limits_{{\it i}=1}^{{\it n}} {\it \eta }_{{\it i}} } {\it y}_{\left(\mathop{\sum }\limits_{{\it i}=1}^{{\it n}} {\it \eta }_{{\it i}} \right)crt} } $ & $\frac{\sqrt{\lambda _{\left(\mathop{\sum }\nolimits_{{\it i}=1}^{{\it n}} {\it \eta }_{{\it i}} \right)} ^{2} +4} -\lambda _{\left(\mathop{\sum }\nolimits_{{\it i}=1}^{{\it n}} {\it \xi }_{{\it i}} \right)} }{2} $ & $\mathop{\sum }\limits_{j>0}^{N=2} \gamma _{j} $ \\
V & $-$1 & $\frac{1}{\lambda _{\left(\mathop{\sum }\nolimits_{{\it i}=1}^{{\it n}} {\it \xi }_{{\it i}} \right)} } $ & 0 & 0 & $\mathop{\sum }\limits_{j=1} \gamma _{j} $ \\
VI & 1 & $\frac{\lambda _{\left(\mathop{\sum }\nolimits_{{\it i}=1}^{{\it n}} {\it \eta }_{{\it i}} \right)} ^{2} {\it \; y}_{\left(\mathop{\sum }\nolimits_{{\it i}=1}^{{\it n}} {\it \eta }_{{\it i}} \right)crt} ^{2} }{\lambda _{\left(\mathop{\sum }\nolimits_{{\it i}=1}^{{\it n}} {\it \xi }_{{\it i}} \right)} } $ & $-\sqrt{{\it \lambda }_{\mathop{\sum }\limits_{{\it i}=1}^{{\it n}} {\it \eta }_{{\it i}} } {\it y}_{\left(\mathop{\sum }\limits_{{\it i}=1}^{{\it n}} {\it \eta }_{{\it i}} \right)crt} } $ & $\frac{\sqrt{\lambda _{\left(\mathop{\sum }\nolimits_{{\it i}=1}^{{\it n}} {\it \eta }_{{\it i}} \right)} ^{2} +4} -\lambda _{\left(\mathop{\sum }\nolimits_{{\it i}=1}^{{\it n}} {\it \xi }_{{\it i}} \right)} }{2} $ & $\mathop{\sum }\limits_{j=1} \gamma _{j} $ \\
VII & $\sqrt{{\it \gamma }} $ & $\frac{\gamma }{\lambda _{\left(\mathop{\sum }\nolimits_{{\it i}=1}^{{\it n}} {\it \xi }_{{\it i}} \right)} } $ & 0 & 0 & $\begin{array}{*{20}{l}}
{\frac{{\psi \sqrt {{\psi ^2} + 4}  - {\psi ^2}}}{2} > \gamma }\\
{where}\\
{\psi  = {\lambda _{\left( {\sum\nolimits_{i = 1}^n {{\xi _i}} } \right)}}}
\end{array}$ \\
VIII & $\sqrt{{\it \beta }_{\mathop{\sum }\limits_{{\it i}=1}^{{\it n}} {\it \xi }_{{\it i}} } {\it y}_{\left(\mathop{\sum }\limits_{{\it i}=1}^{{\it n}} {\it \xi }_{{\it i}} \right)crt} } $ & $\frac{\sqrt{\lambda _{\left(\mathop{\sum }\nolimits_{{\it i}=1}^{{\it n}} {\it \xi }_{{\it i}} \right)} ^{2} +4} -\lambda _{\left(\mathop{\sum }\nolimits_{{\it i}=1}^{{\it n}} {\it \xi }_{{\it i}} \right)} }{2} $ & 0 & 0 & $\mathop{\sum }\limits_{j>0}^{N=2} \gamma _{j} $ \\ \hline \hline
\end{tabular}
\end{table*}
Now from Eq. (\ref{1a29}), we determine the solutions of the self-similar nature, that is
\be \begin{array}{l}
\frac{{\dot H}}{{{H^2}}} = \frac{{HH'}}{{{H^2}}} = \frac{{H'}}{H}\\
 =  - \frac{3}{2}\left( { - \frac{{\;{y_{\sum _{i = 1}^n{\xi _i}}}\left( {\gamma  - \;x_{\sum _{i = 1}^n{\xi _i}}^2} \right)}}{{\sqrt {1 - \sum _{i = 1}^n\dot \xi _i^2} }} - \frac{{\;{y_{\sum _{i = 1}^n{\eta _i}}}(\gamma  - \;x_{\sum _{i = 1}^n{\eta _i}}^2}}{{\sqrt {1 - \sum _{i = 1}^n\dot \eta _i^2} }} + \gamma } \right)
\end{array} \label{1a43}\ee
This corresponds to an expanding universe such that $a\left(t\right)$, the scale factor scales like $a\left(t\right)\propto \; t^{p} $, where
\be p=\frac{2}{3\left(-\frac{\; y_{\mathop{\sum }\nolimits_{i=1}^{n} \xi _{i} } \left(\gamma -\; x_{\mathop{\sum }\nolimits_{i=1}^{n} \xi _{i} } ^{2} \right)}{\sqrt{1-\mathop{\sum }\nolimits_{i=1}^{n} \dot{\xi }_{i} ^{2} } } -\frac{\; y_{\mathop{\sum }\nolimits_{i=1}^{n} \eta _{i} } (\gamma -\; x_{\mathop{\sum }\nolimits_{i=1}^{n} \eta _{i} } ^{2} }{\sqrt{1-\mathop{\sum }\nolimits_{i=1}^{n} \dot{\eta }_{i} ^{2} } } +\gamma \right)} \label{1a44}\ee
We now study the stability around the critical points given in Table \ref{Tab:tab1} for which we consider small perturbations $\; \delta x_{\mathop{\sum }\limits_{i=1}^{n} \xi _{i} } ,\; \; \delta y_{\mathop{\sum }\limits_{i=1}^{n} \xi _{i} } ,\; \; \delta x_{\mathop{\sum }\limits_{i=1}^{n} \eta _{i} } ,\; \; \delta y_{\mathop{\sum }\limits_{i=1}^{n} \eta _{i} } \; $about the critical points  $x_{\mathop{\sum }\limits_{i=1}^{n} \xi _{i} } ,\; \; y_{\mathop{\sum }\limits_{i=1}^{n} \xi _{i} } ,\; \; x_{\mathop{\sum }\limits_{i=1}^{n} \eta _{i} } ,\; \; y_{\mathop{\sum }\limits_{i=1}^{n} \eta _{i} } $ respectively such that
\[x_{\mathop{\sum }\limits_{i=1}^{n} \xi _{i} } \to x_{\mathop{\sum }\limits_{i=1}^{n} \xi _{i} } +\delta x_{\mathop{\sum }\limits_{i=1}^{n} \xi _{i} } \]
\[\; y_{\mathop{\sum }\limits_{i=1}^{n} \xi _{i} } \to y_{\mathop{\sum }\limits_{i=1}^{n} \xi _{i} } +\delta y_{\mathop{\sum }\limits_{i=1}^{n} \xi _{i} } \]
\[\; x_{\mathop{\sum }\limits_{i=1}^{n} \eta _{i} } \to x_{\mathop{\sum }\limits_{i=1}^{n} \eta _{i} } +\delta x_{\mathop{\sum }\limits_{i=1}^{n} \eta _{i} } \]
\[\; y_{\mathop{\sum }\limits_{i=1}^{n} \eta _{i} } \to \; y_{\mathop{\sum }\limits_{i=1}^{n} \eta _{i} } +\; \delta y_{\mathop{\sum }\limits_{i=1}^{n} \eta _{i} } \]
When we substitute these in Eq. (\ref{1a37}), Eq. (\ref{1a38}), Eq. (\ref{1a40}) and Eq. (\ref{1a41}), these equations lead to the following equation in matrix form which represents differential equations of the first order.
\be \left(\begin{array}{c} {\delta x'_{\mathop{\sum }\limits_{i=1}^{n} \xi _{i} } } \\ {\delta y'_{\mathop{\sum }\limits_{i=1}^{n} \xi _{i} } } \\ {\delta x'_{\mathop{\sum }\limits_{i=1}^{n} \eta _{i} } } \\ {\delta y'_{\mathop{\sum }\limits_{i=1}^{n} \eta _{i} } } \end{array}\right)=X\left(\begin{array}{c} {\delta x_{\mathop{\sum }\limits_{i=1}^{n} \xi _{i} } } \\ {\delta y_{\mathop{\sum }\limits_{i=1}^{n} \xi _{i} } } \\ {\delta x_{\mathop{\sum }\limits_{i=1}^{n} \eta _{i} } } \\ {\; \delta y_{\mathop{\sum }\limits_{i=1}^{n} \eta _{i} } } \end{array}\right)\label{1a45}\ee
The dependence of the matrix $X$ on ${\it x}_{\left(\mathop{\sum }\limits_{{\it i}=1}^{{\it n}} {\it \xi }_{{\it i}} \right)crt} $ ${\it y}_{\left(\mathop{\sum }\limits_{{\it i}=1}^{{\it n}} {\it \xi }_{{\it i}} \right)crt} $ ${\it x}_{\left(\mathop{\sum }\limits_{{\it i}=1}^{{\it n}} {\it \eta }_{{\it i}} \right)crt} $ and ${\it \; y}_{\left(\mathop{\sum }\limits_{{\it i}=1}^{{\it n}} {\it \eta }_{{\it i}} \right)crt} $ is clear. Now the general solution for the evolution of linear perturbations can be expressed in the following way using the eigenvalues a, b, c, and d of the matrix $X$.
\be \delta x_{\mathop{\sum }\limits_{i=1}^{n} \xi _{i} } =v_{11} e^{aN} +v_{12} e^{bN} +v_{13} e^{cN} +v_{14} e^{dN} \label{1a46}\ee
\be \delta y_{\mathop{\sum }\limits_{i=1}^{n} \xi _{i} } =v_{21} e^{aN} +v_{22} e^{bN} +v_{23} e^{cN} +v_{24} e^{dN} \label{1a47}\ee
\be \delta x_{\mathop{\sum }\limits_{i=1}^{n} \eta _{i} } =v_{31} e^{aN} +v_{32} e^{bN} +v_{33} e^{cN} +v_{34} e^{dN} \label{1a48}\ee
\be \; \delta y_{\mathop{\sum }\limits_{i=1}^{n} \eta _{i} } =v_{41} e^{aN} +v_{42} e^{bN} +v_{43} e^{cN} +v_{44} e^{dN} \label{1a49}\ee

\begin{table*}[ht!]
\centering
\caption{Enlisting the eigenvalues and status of stability}\label{Table:Tab2}
  \begin{tabular}{cccccc}
\hline
\hline
\\
\textbf{Labels} & \textbf{A} & \textbf{B} & \textbf{C} & \textbf{D} & \textbf{Status of Stability} \\ \\ \hline
\\ I &  $-3$ &  $3\gamma $ &  $-3$ &  $3\gamma $ & unstable \\
\\ II &  $-3$ &  $-3{\it x}_{\left(\mathop{\sum }\limits_{{\it i}=1}^{{\it n}} {\it \eta }_{{\it i}} \right)crt} ^{2} $ &  $-\frac{3}{2} \left(2+{\it x}_{\left(\mathop{\sum }\limits_{{\it i}=1}^{{\it n}} {\it \eta }_{{\it i}} \right)crt} ^{2} \right)$ &  $-3\gamma \left(\gamma +{\it x}_{\left(\mathop{\sum }\limits_{{\it i}=1}^{{\it n}} {\it \eta }_{{\it i}} \right)crt} ^{2} \right)$ & stable \\
\\ III &  $6$ &  $3\gamma $ &  $-3$ &  $3\gamma $ & unstable \\
\\ IV &  $6$ & $-3{\it x}_{\left(\mathop{\sum }\limits_{{\it i}=1}^{{\it n}} {\it \eta }_{{\it i}} \right)crt} ^{2} $ &  $-\frac{3}{2} \left(2+{\it x}_{\left(\mathop{\sum }\limits_{{\it i}=1}^{{\it n}} {\it \eta }_{{\it i}} \right)crt} ^{2} \right)$ &  $-3\gamma \left(\gamma +{\it x}_{\left(\mathop{\sum }\limits_{{\it i}=1}^{{\it n}} {\it \eta }_{{\it i}} \right)crt} ^{2} \right)$ & unstable \\
\\ V &  $0$ &  $-\frac{3}{2} $ &  $-3$ &  $3$ & unstable \\
\\ VI &  $6+6{\it x}_{\left(\mathop{\sum }\limits_{{\it i}=1}^{{\it n}} {\it \eta \; }_{{\it i}} \right)crt} ^{2} $ &  $\frac{3{\it x}_{\left(\mathop{\sum }\nolimits_{{\it i}=1}^{{\it n}} {\it \eta }_{{\it i}} \right)crt} ^{2} }{2} $ &  $-\frac{3}{2} \left(2+{\it x}_{\left(\mathop{\sum }\limits_{{\it i}=1}^{{\it n}} {\it \eta }_{{\it i}} \right)crt} ^{2} \right)$ &  $-3\gamma \left(\gamma +{\it x}_{\left(\mathop{\sum }\limits_{{\it i}=1}^{{\it n}} {\it \eta }_{{\it i}} \right)crt} ^{2} \right)$ & unstable \\
\\ VII &  $h$ &  $k$ &  $-3$ &  $3\gamma $ & unstable \\
\\ VIII &  $-3\gamma +3{\it x}_{\left(\mathop{\sum }\limits_{{\it i}=1}^{{\it n}} {\it \xi }_{{\it i}} \right)crt} ^{2} $ &  $-3+\frac{3{\it x}_{\left(\mathop{\sum }\nolimits_{{\it i}=1}^{{\it n}} {\it \xi }_{{\it i}} \right)crt} ^{2} }{2} $ &  $-3$ &  $3{\it x}_{\left(\mathop{\sum }\limits_{{\it i}=1}^{{\it n}} {\it \xi }_{{\it i}} \right)crt} ^{2} $ & unstable \\ \\
\hline \hline
\end{tabular}
\end{table*}

\be
\begin{tiny} h=\frac{3\left[\lambda _{\left(\mathop{\sum }\nolimits_{{\it i}=1}^{{\it n}} {\it \xi }_{{\it i}} \right)} \left(\gamma -2\right)\mp \sqrt{16\lambda _{\left(\mathop{\sum }\nolimits_{{\it i}=1}^{{\it n}} {\it \xi }_{{\it i}} \right)} \gamma ^{2} sqrt\left(1-\gamma \right)+\lambda _{\left(\mathop{\sum }\nolimits_{{\it i}=1}^{{\it n}} {\it \xi }_{{\it i}} \right)}^{2} \left(4-20\gamma +17\gamma ^{2} \right)} \right]}{4\lambda _{\left(\mathop{\sum }\nolimits_{{\it i}=1}^{{\it n}} {\it \xi }_{{\it i}} \right)}} \end{tiny} \label{1a50}\ee
and
\be \begin{tiny}
k=\frac{3\left[\lambda _{\left(\mathop{\sum }\nolimits_{{\it i}=1}^{{\it n}} {\it \xi }_{{\it i}} \right)} \left(\gamma -2\right)\mp \sqrt{16\lambda _{\left(\mathop{\sum }\nolimits_{{\it i}=1}^{{\it n}} {\it \xi }_{{\it i}} \right)} \gamma ^{2} sqrt\left(1-\gamma \right)+\lambda _{\left(\mathop{\sum }\nolimits_{{\it i}=1}^{{\it n}} {\it \xi }_{{\it i}} \right)}^{2} \left(4-20\gamma +17\gamma ^{2} \right)} \right]}{4\lambda _{\left(\mathop{\sum }\nolimits_{{\it i}=1}^{{\it n}} {\it \xi }_{{\it i}} \right)}} \end{tiny} \label{1a51}\ee
E.J. Copeland et al. \cite{Pd26} and Z.K. Guo et al. \cite{Pd27} have shown that the real parts of all eigenvalues might be negative for the remaining stable points. Table \ref{Table:Tab2} lists all the eigenvalues and status of their stability. A fixed critical point I gives a solution with fluid domination, the points II and III represent a phantom tachyon and kinetical tachyon domination in solutions respectively. A two- field dominated solution is manifested by the fixed critical point IV. For $\gamma $ unity the fixed critical points V and VI have their existence. The energy densities $\rho _{\left(\mathop{\sum }\limits_{{\it i}=1}^{{\it n}} {\it \xi }_{{\it i}} \right)} $ and $\rho _{\left(\gamma \right)} $ show a decrease at same rate in the point VII when the point VIII indicate a solution where tachyon field energy dominates.

\begin{figure*}[ht!]
\centering
\begin{center}
 \includegraphics[scale=0.50]{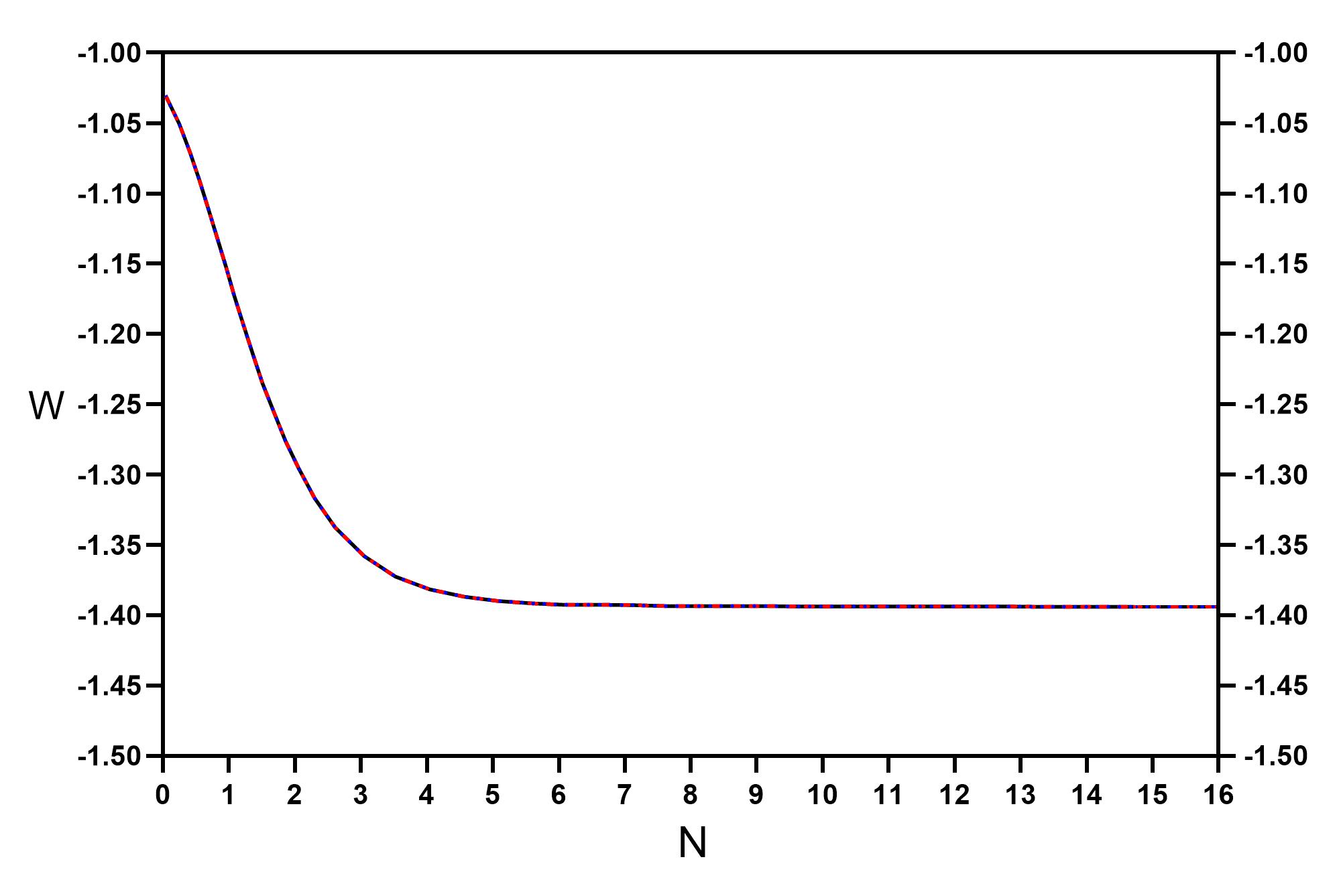}
 \includegraphics[scale=0.50]{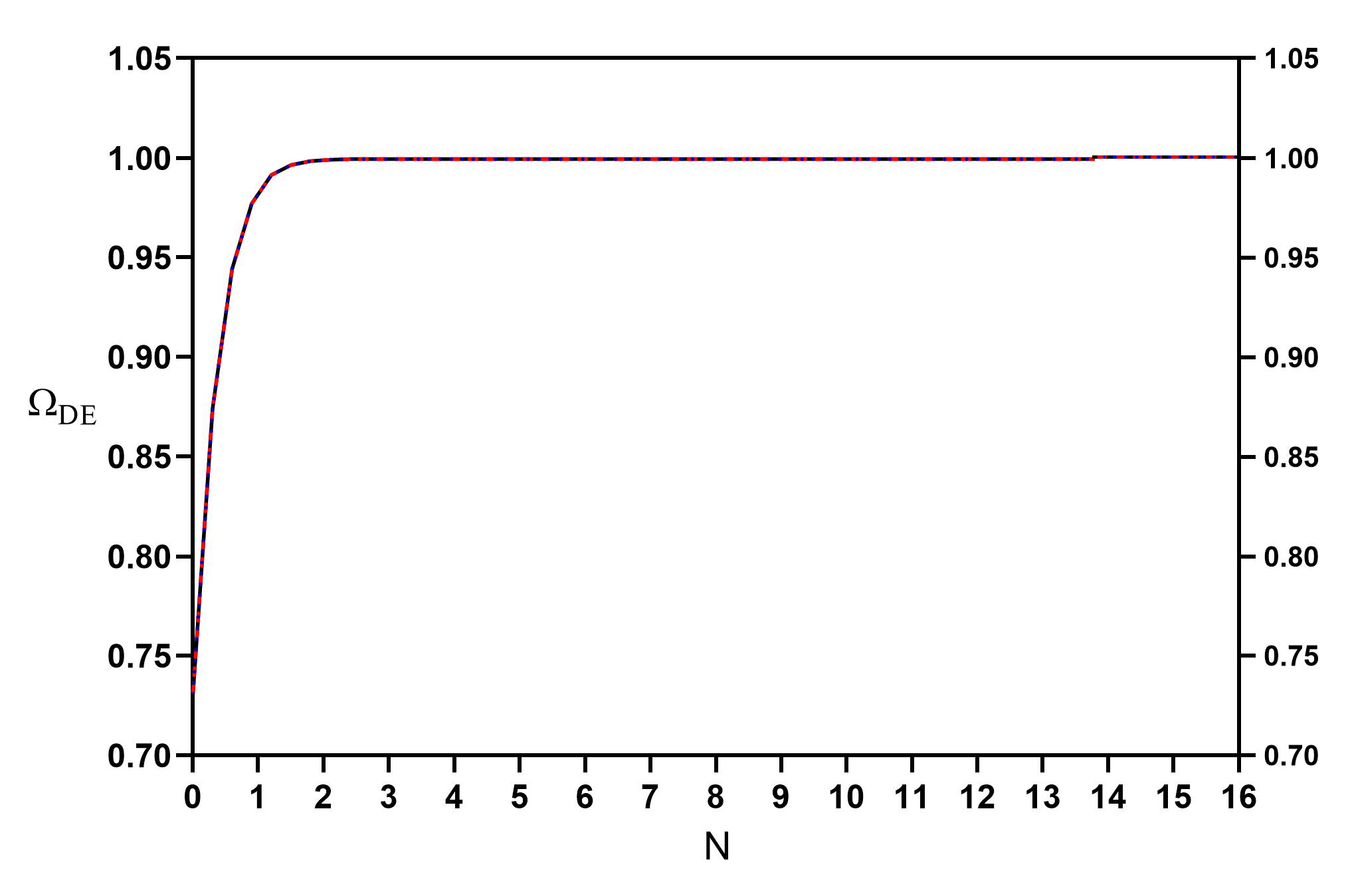}
 \includegraphics[scale=0.50]{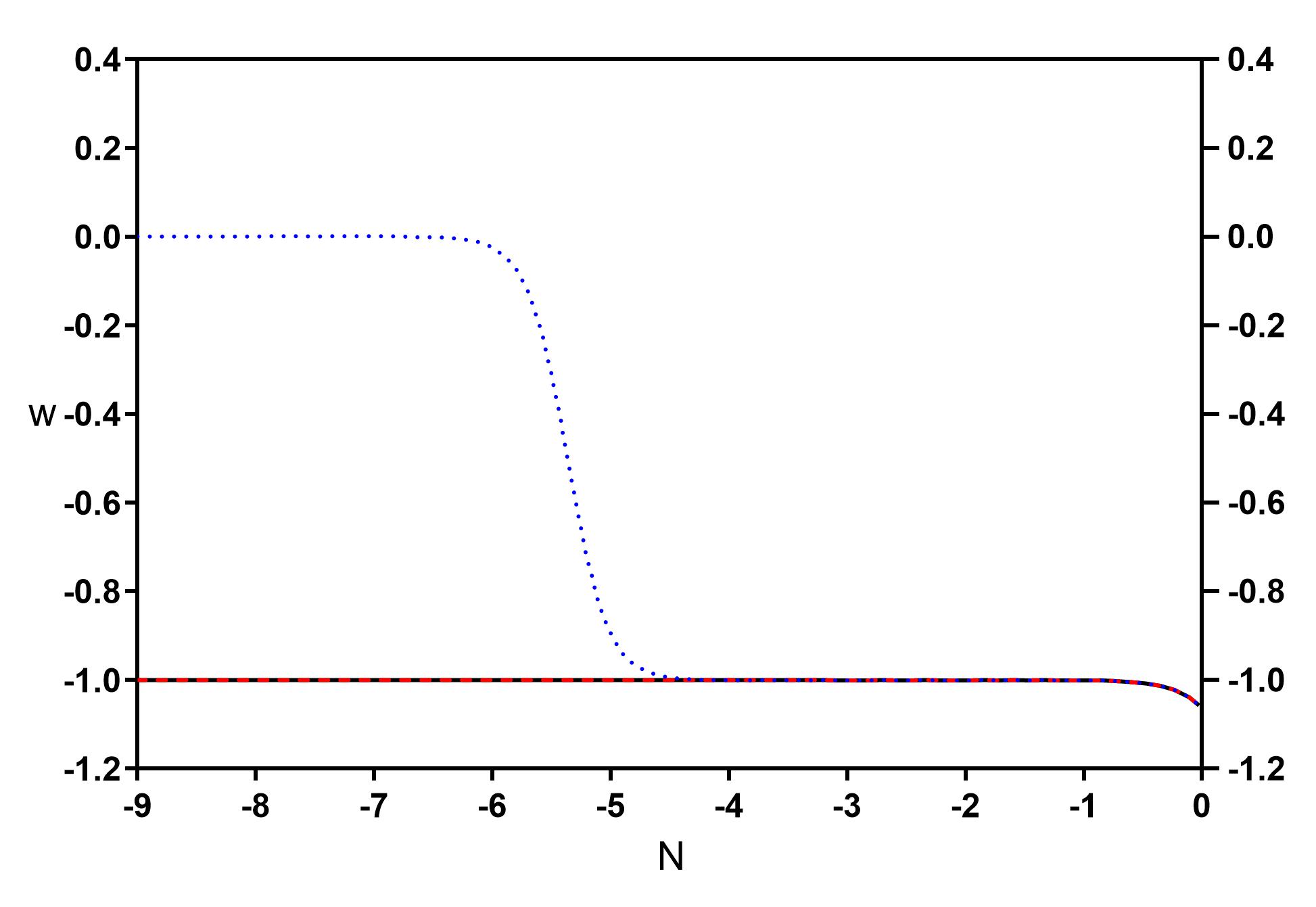}
 \includegraphics[scale=0.50]{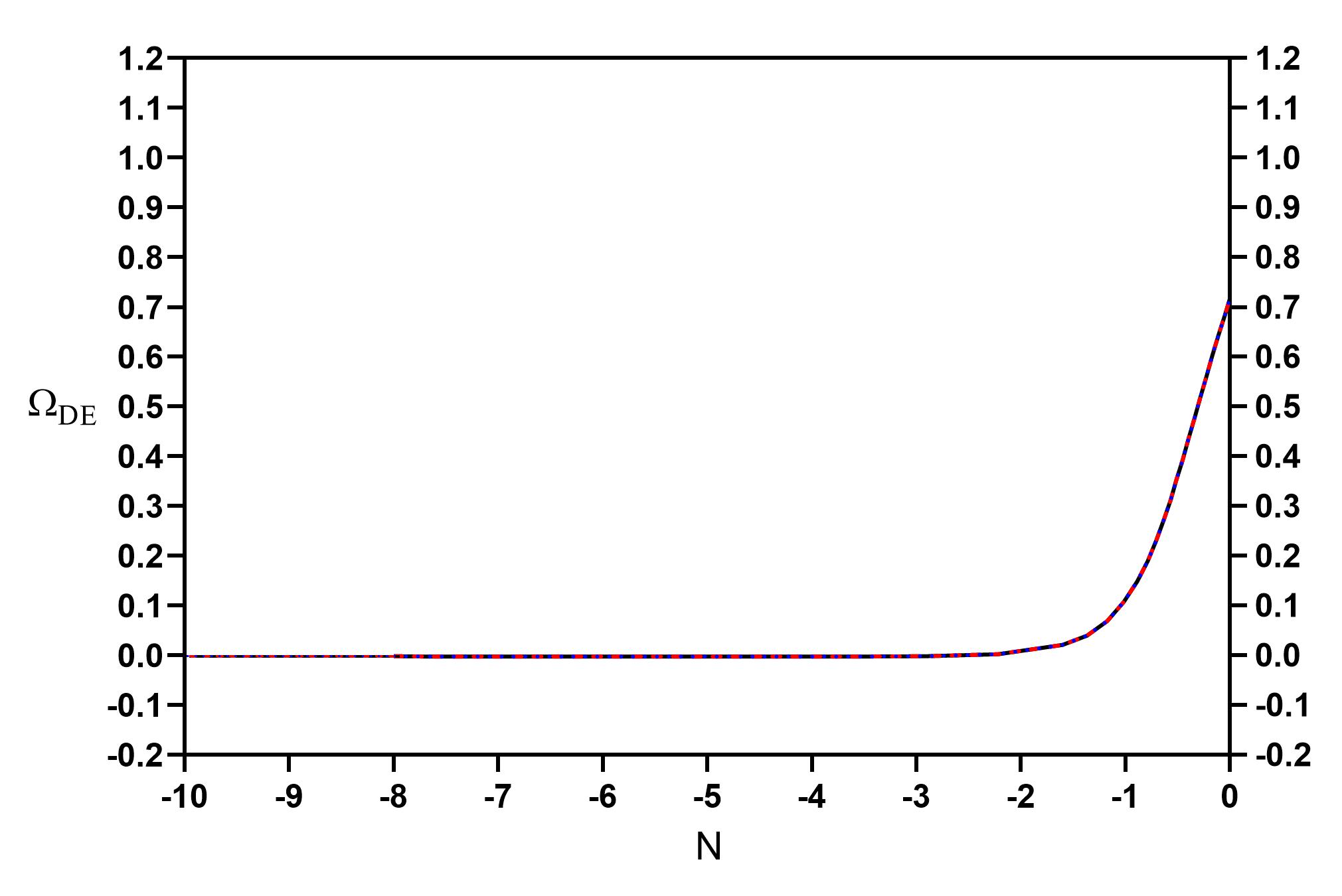}
    \caption{The diagrams demonstrating the evolution of equation of state parameter $w$ and the development of the parameter of dark energy ${\it }_{DE}$ in the very early Universe and their development during its behavior in late time accelerated expansion where $\gamma =1$ and ${{\beta } {\sum\limits_{i=1}^{n}{{{\xi }_{i}}}}}$ and ${{\beta }_{\sum\limits_{i=1}^{n}{{{\eta }_{i}}}}}$ both take on the value 0.33.} \label{Figure: Fig3}
\end{center}
\end{figure*}
\pagebreak
The above diagrams in Fig.\ref{Figure: Fig3} illustrate the plot of the evolution of the equation of state parameter $w$ and dark energy parameter ${\it }_{DE} $  cosmologically against the growth of e-folding number $N$, It can be seen that the evolutionary path cosmologically bends towards the point II which is a fixed stability point in the Table \ref{Tab:tab1} in the model. A comparison between the plots of the Fig.\ref{Figure: Fig1} and the values of the critical points listed in Table \ref{Tab:tab1}, can clarify the situation further.

When we plug  $\frac{1}{3} $  the value of $\lambda _{\left(\mathop{\sum }\limits_{{\it i}=1}^{{\it n}} {\it \eta }_{{\it i}} \right)} $ in point II the stable critical point i.e.
${\it x}_{\left(\mathop{\sum }\limits_{{\it i}=1}^{{\it n}} {\it \xi }_{{\it i}} \right)crt} =0$ , ${\it y}_{\left(\mathop{\sum }\limits_{{\it i}=1}^{{\it n}} {\it \xi }_{{\it i}} \right)crt} =0{\rm \; }$, ${\it x}_{\left(\mathop{\sum }\limits_{{\it i}=1}^{{\it n}} {\it \eta }_{{\it i}} \right)crt} =-\sqrt{{\it \lambda }_{\mathop{\sum }\limits_{{\it i}=1}^{{\it n}} {\it \eta }_{{\it i}} } {\it y}_{\left(\mathop{\sum }\limits_{{\it i}=1}^{{\it n}} {\it \eta }_{{\it i}} \right)crt} } $
and
\be{\it \; y}_{\left(\mathop{\sum }\limits_{{\it i}=1}^{{\it n}} {\it \eta }_{{\it i}} \right)crt} =\frac{\sqrt{\lambda _{\left(\mathop{\sum }\nolimits_{{\it i}=1}^{{\it n}} {\it \eta }_{{\it i}} \right)} ^{2} +4} +\lambda _{\left(\mathop{\sum }\nolimits_{{\it i}=1}^{{\it n}} {\it \eta }_{{\it i}} \right)} }{2} \label{1a52}\ee
we obtain the values of these points
${\it x}_{\left(\mathop{\sum }\limits_{{\it i}=1}^{{\it n}} {\it \xi }_{{\it i}} \right)crt} =0$ , ${\it y}_{\left(\mathop{\sum }\limits_{{\it i}=1}^{{\it n}} {\it \xi }_{{\it i}} \right)crt} =0{\rm \; }$, ${\it x}_{\left(\mathop{\sum }\limits_{{\it i}=1}^{{\it n}} {\it \eta }_{{\it i}} \right)crt} =-\sqrt{{\it \lambda }_{\mathop{\sum }\limits_{{\it i}=1}^{{\it n}} {\it \eta }_{{\it i}} } {\it y}_{\left(\mathop{\sum }\limits_{{\it i}=1}^{{\it n}} {\it \eta }_{{\it i}} \right)crt} } =-0.627285$ and
\be {\it \; y}_{\left(\mathop{\sum }\limits_{{\it i}=1}^{{\it n}} {\it \eta }_{{\it i}} \right)crt} =\frac{\sqrt{\lambda _{\left(\mathop{\sum }\nolimits_{{\it i}=1}^{{\it n}} {\it \eta }_{{\it i}} \right)} ^{2} +4} +\lambda _{\left(\mathop{\sum }\nolimits_{{\it i}=1}^{{\it n}} {\it \eta }_{{\it i}} \right)} }{2} =1.180460 \label{1a53}\ee
which show consistency with plotting in Fig.\ref{Figure: Fig1}. Further with the help of Eq.    G at the fixed point II, we possess $w=\frac{p}{\rho } =-1-{\it x}_{\left(\mathop{\sum }\limits_{{\it i}=1}^{{\it n}} {\it \eta }_{{\it i}} \right)crt} ^{2} =-1.393487$ which also shows consistency with the plotting drawn in Fig.\ref{Figure: Fig3}.
In Table \ref{Tab:tab1}, We would like to describe the initial values of the points $x_{\mathop{\sum }\limits_{i=1}^{n} \xi _{i} } ,\; \; y_{\mathop{\sum }\limits_{i=1}^{n} \xi _{i} } ,\; \; x_{\mathop{\sum }\limits_{i=1}^{n} \eta _{i} } ,\; $and  $\; y_{\mathop{\sum }\limits_{i=1}^{n} \eta _{i} } $ would evolve towards stability If these are not the values of unstable points in the model granted the condition $1-{\it x}_{\left(\mathop{\sum }\limits_{{\it i}=1}^{{\it n}} {\it \xi }_{{\it i}} \right)} ^{2} >0$ does not get violated as physical constraint. When the values of the points  $x_{\mathop{\sum }\limits_{i=1}^{n} \xi _{i} } ,\; \; y_{\mathop{\sum }\limits_{i=1}^{n} \xi _{i} } ,\; \; x_{\mathop{\sum }\limits_{i=1}^{n} \eta _{i} } ,\; $and  $\; y_{\mathop{\sum }\limits_{i=1}^{n} \eta _{i} } $ deviate slightly by a quantity amounting $\vec{\delta }$ from the values of  ${\it x}_{\left(\mathop{\sum }\limits_{{\it i}=1}^{{\it n}} {\it \xi }_{{\it i}} \right)crt} ,$ ${\it y}_{\left(\mathop{\sum }\limits_{{\it i}=1}^{{\it n}} {\it \xi }_{{\it i}} \right)crt} $,  ${\it x}_{\left(\mathop{\sum }\limits_{{\it i}=1}^{{\it n}} {\it \eta }_{{\it i}} \right)crt} ,$ ${\it \; y}_{\left(\mathop{\sum }\limits_{{\it i}=1}^{{\it n}} {\it \eta }_{{\it i}} \right)crt} ,$ from the Eq. (\ref{53}), It can be envisaged that $\vec{\delta }$ may be larger despite diminishing.

\section{Final remarks and summary}
The analysis shows that the model does not indicate sensitivity to the kinetic energy density of under consideration multi-fields initially. It has been shown that there exists a stable unique critical point during the analysis of background spatially flat universe in the phase space. We make its comparison with the tachyon model altogether.
 $M_{\mathop{\sum }\limits_{i=1}^{n} \xi _{i} } $ should be larger \cite{Pd28} enough than $M_{PL} $ in case of dark energy of multi-field tachyon with inverse square potential of the type $V\left(\mathop{\sum }\limits_{{\it i}=1}^{{\it n}} {\it \xi }_{{\it i}} \right)=M_{\mathop{\sum }\limits_{i=1}^{n} \xi _{i} }^{2} \mathop{\sum }\limits_{i=1}^{n} \xi _{i} ^{-2}$ in order to meet the late time acceleration of the cosmos $i.e.$ $a\propto \left[time\right]^{\frac{1}{2} \left(\frac{M_{\mathop{\sum }\nolimits_{i=1}^{n} \xi _{i} } }{M_{PL} } \right)^{2} \gg 1}$. This huge mass pushes the solutions towards dense energy regions where even General Theory of Relativity fails, therefore the potential $V\left(\mathop{\sum }\limits_{{\it i}=1}^{{\it n}} {\it \xi }_{{\it i}} \right)=M_{\mathop{\sum }\limits_{i=1}^{n} \xi _{i} }^{4-p} \mathop{\sum }\limits_{i=1}^{n} \xi _{i} ^{-p} $ where $p$ lies between $0$ and $2$. The phantom tachyon fields $\mathop{\sum }\limits_{{\it i}=1}^{{\it n}} {\it \eta }_{{\it i}} $ in our model takes the responsibility of this late time acceleration with equation of state parameter $w_{\mathop{\sum }\limits_{{\it i}=1}^{{\it n}} {\it \eta }_{{\it i}} } <-1$. From Eq. (\ref{1a37}) and Eq. (\ref{1a38}), it becomes clear that the value of $M_{\mathop{\sum }\limits_{i=1}^{n} \xi _{i} } $ is not still smaller as required by recent observational constraints. This is due to the reason for $\lambda _{M_{\mathop{\sum }\limits_{i=1}^{n} \xi _{i} } } =\frac{4}{3} \frac{M_{PL}^{2} }{M_{\mathop{\sum }\nolimits_{i=1}^{n} \xi _{i} }^{2} } $ being larger, the factor $1-\left(\mathop{\sum }\limits_{i=1}^{n} \dot{\xi }_{i} \right)^{2} $falls in risk of non-positive behavioral increment.
  In our multi-fields model of tachyon and phantom tachyon there is only one stable critical point namely II whose value does not rest on the value of $\gamma $, on the other hand in reference to \cite{Pd29}, it is shown that in tachyon model the sole source of dark energy is tachyon and it has three critical stable points whose existence hinges upon $\gamma $. Further we have seen the values of scalar fields  $\mathop{\sum }\limits_{i=1}^{n} \dot{\xi }_{i} $ and  $\mathop{\sum }\limits_{i=1}^{n} \dot{\eta }_{i} $ are zero at critical points in our model of tachyon and phantom tachyon but in tachyon singly model of dark energy the values of both fields at critical points can be non-zero also as depicted in the Table [1] The values of   ${\it x}_{\left(\mathop{\sum }\limits_{{\it i}=1}^{{\it n}} {\it \xi }_{{\it i}} \right)crt} $, ${\it y}_{\left(\mathop{\sum }\limits_{{\it i}=1}^{{\it n}} {\it \xi }_{{\it i}} \right)crt} $, ${\it x}_{\left(\mathop{\sum }\limits_{{\it i}=1}^{{\it n}} {\it \eta }_{{\it i}} \right)crt} $  and ${\it \; y}_{\left(\mathop{\sum }\limits_{{\it i}=1}^{{\it n}} {\it \eta }_{{\it i}} \right)crt} $ at the critical points are fixed with the value of ${\it x}_{\left(\mathop{\sum }\limits_{{\it i}=1}^{{\it n}} {\it \eta }_{{\it i}} \right)crt} $ becoming non-zero and if its value is zero then ${\it \; y}_{\left(\mathop{\sum }\limits_{{\it i}=1}^{{\it n}} {\it \eta }_{{\it i}} \right)crt} $ turns out to be zero, this is obvious from the Eq. [1.40] , it further ensue the impossibility that ${\it x}_{\left(\mathop{\sum }\limits_{{\it i}=1}^{{\it n}} {\it \eta }_{{\it i}} \right)crt} $  and  ${\it \rho }_{\left(\mathop{\sum }\limits_{{\it i}=1}^{{\it n}} {\it \eta }_{{\it i}} \right)crt} $ are zero $i.e.$
  \be \begin{array}{c}
{\rho _{\left( {\mathop \sum \limits_{i = 1}^n {\xi _i}} \right)crt}} = \frac{{V\left[ {\left( {\mathop \sum \nolimits_{i = 1}^n {\xi _i}} \right)crt} \right]}}{{\sqrt {1 - \left( {\mathop \sum \nolimits_{i = 1}^n {{\dot \xi }_i}} \right)cr{t^2}} }} = \frac{{3M_{PL}^2\;{y_{\left( {\mathop \sum \nolimits_{i = 1}^n {\xi _i}} \right)crt}}{H^2}}}{{\sqrt {1 - {x_{\left( {\mathop \sum \nolimits_{i = 1}^n {\xi _i}} \right)crt}}^2} }}\\
{p_{\left( {\mathop \sum \limits_{i = 1}^n {\xi _i}} \right)crt}} = \left( {{w_{\left( {\mathop \sum \limits_{i = 1}^n {\xi _i}} \right)crt}}} \right)\left( {{\rho _{\left( {\mathop \sum \limits_{i = 1}^n {\xi _i}} \right)crt}}} \right)\\
\Rightarrow {w_{\left( {\mathop \sum \limits_{i = 1}^n {\xi _i}} \right)crt}} = 1 - \left( {\mathop \sum \limits_{i = 1}^n {{\dot \xi }_i}} \right)cr{t^2} = 1 - {x_{\left( {\mathop \sum \limits_{i = 1}^n {\xi _i}} \right)crt}}^2 \ge  - 1
\end{array}\label{1a54}\ee
and
\be \begin{array}{c}
{\rho _{\left( {\mathop \sum \limits_{i = 1}^n {\eta _i}} \right)crt}} = \frac{{V\left[ {\left( {\mathop \sum \nolimits_{i = 1}^n {\eta _i}} \right)crt} \right]}}{{\sqrt {1 - \left( {\mathop \sum \nolimits_{i = 1}^n {{\dot \eta }_i}} \right)cr{t^2}} }} = \frac{{3M_{PL}^2\;{y_{\left( {\mathop \sum \nolimits_{i = 1}^n {\eta _i}} \right)crt}}{H^2}}}{{\sqrt {1 - {x_{\left( {\mathop \sum \nolimits_{i = 1}^n {\eta _i}} \right)crt}}^2} }}\\
{p_{\left( {\mathop \sum \limits_{i = 1}^n {\eta _i}} \right)crt}} = \left( {{w_{\left( {\mathop \sum \limits_{i = 1}^n {\eta _i}} \right)crt}}} \right)\left( {{\rho _{\left( {\mathop \sum \limits_{i = 1}^n {\eta _i}} \right)crt}}} \right)\\
 \Rightarrow {w_{\left( {\mathop \sum \limits_{i = 1}^n {\eta _i}} \right)crt}} = 1 - \left( {\mathop \sum \limits_{i = 1}^n {{\dot \eta }_i}} \right)cr{t^2} = 1 - {x_{\left( {\mathop \sum \limits_{i = 1}^n {\eta _i}} \right)crt}}^2 <  - 1
\end{array}\label{1a55}\ee
It is clear from Eq. (\ref{1a54}) that $H^{2} $ becomes non-increasing for non-zero ${\it \; y}_{\left(\mathop{\sum }\limits_{{\it i}=1}^{{\it n}} {\it \xi }_{{\it i}} \right)crt} $ at the points which are fixed, with ${\it \rho }_{\left(\mathop{\sum }\limits_{{\it i}=1}^{{\it n}} {\it \xi }_{{\it i}} \right)crt} $ also non-increasing. The same is not inapplicable from Eq. (\ref{1a55}) for ${\it \; y}_{\left(\mathop{\sum }\limits_{{\it i}=1}^{{\it n}} {\it \eta }_{{\it i}} \right)crt}$ and ${\it \rho }_{\left(\mathop{\sum }\limits_{{\it i}=1}^{{\it n}} {\it \eta }_{{\it i}} \right)crt}$  making one of the two critical points ${\it \; y}_{\left(\mathop{\sum }\limits_{{\it i}=1}^{{\it n}} {\it \xi }_{{\it i}} \right)crt} $, ${\it \; y}_{\left(\mathop{\sum }\limits_{{\it i}=1}^{{\it n}} {\it \eta }_{{\it i}} \right)crt} $  equal to zero. Therefore ${\it \; y}_{\left(\mathop{\sum }\limits_{{\it i}=1}^{{\it n}} {\it \xi }_{{\it i}} \right)crt} $  is set to zero for ${\it \rho }_{\left(\mathop{\sum }\limits_{{\it i}=1}^{{\it n}} {\it \xi }_{{\it i}} \right)crt} $  as non-increasing and ${\it \rho }_{\left(\mathop{\sum }\limits_{{\it i}=1}^{{\it n}} {\it \eta }_{{\it i}} \right)crt} $  as increasing. Moreover if $1-{\it x}_{\left(\mathop{\sum }\limits_{{\it i}=1}^{{\it n}} {\it \xi }_{{\it i}} \right)crt} ^{2} >0$  does not get violated and for ${\it \; y}_{\left(\mathop{\sum }\limits_{{\it i}=1}^{{\it n}} {\it \xi }_{{\it i}} \right)crt} $  being zero, the critical point ${\it \; x}_{\left(\mathop{\sum }\limits_{{\it i}=1}^{{\it n}} {\it \xi }_{{\it i}} \right)crt} $  also becomes zero.
In tachyon model of dark energy, the speed of sound is expressed \cite{Pd30} by the expression of the form
Speed of sound \be c_{s}^{2} =\frac{{\it p}_{\left(\mathop{\sum }\nolimits_{{\it i}=1}^{{\it n}} {\it \xi }_{{\it i}} \right)\frac{1}{2} \partial _{\mu } \left(\mathop{\sum }\nolimits_{{\it i}=1}^{{\it n}} {\it \xi }_{{\it i}} \right)^{2} } }{{\it \rho }_{\left(\mathop{\sum }\nolimits_{{\it i}=1}^{{\it n}} {\it \xi }_{{\it i}} \right)\frac{1}{2} \partial _{\mu } \left(\mathop{\sum }\nolimits_{{\it i}=1}^{{\it n}} {\it \xi }_{{\it i}} \right)^{2} } } =1-\left(\mathop{\sum }\limits_{i=1}^{n} \dot{\xi }_{i} \right)^{2} \label{1a56}\ee
\noindent Due to the under-root present in the Langrangian density, the difference term $1-\left(\mathop{\sum }\limits_{i=1}^{n} \dot{\xi }_{i} \right)^{2} $ will be non-positive accordingly and resultantly it helps pressure and energy remain in the realm of real, therefore a positive sound speed is attributed to homogeneous perturbations which  owes stability. We can put to use independent sound speed of each component of the multi-fields tachyon and phantom tachyon in our case here to give a description to the model. It is, however, notable that J.Q. Xia et al. \cite{Pd31} and H. Kodama et al. \cite{Pd32} have shown the use of effective sound speed because using of two or more independent components of sound does not coincide with the present juncture of the constraints of dark energy. The effective sound speed for larger $N$ in the case when the energy density is considered as a fraction of dark energy density of phantom tachyon, when ${\it }_{\left(\mathop{\sum }\limits_{{\it i}=1}^{{\it n}} {\it \eta }_{{\it i}} \right)} $ tends to unity, is

\be c_{s}^{2} =\frac{{\it p}_{\left(\mathop{\sum }\nolimits_{{\it i}=1}^{{\it n}} {\it \xi }_{{\it i}} \right)\frac{1}{2} \partial _{\mu } \left(\mathop{\sum }\nolimits_{{\it i}=1}^{{\it n}} {\it \xi }_{{\it i}} \right)^{2} } }{{\it \rho }_{\left(\mathop{\sum }\nolimits_{{\it i}=1}^{{\it n}} {\it \xi }_{{\it i}} \right)\frac{1}{2} \partial _{\mu } \left(\mathop{\sum }\nolimits_{{\it i}=1}^{{\it n}} {\it \xi }_{{\it i}} \right)^{2} } } =1+\left(\mathop{\sum }\limits_{i=1}^{n} \dot{\xi }_{i} \right)^{2} >1\label{1a57}\ee
for the effective density in the Langrangian of Eq. (\ref{1a2}). It can be interpreted as the perturbations of the scalar field in the background that can move with a speed larger than the speed of light in the preferential and privileged frame of reference where the field in the background is homogeneously existing. Conclusively we investigated a dark energy model consisting of multi-field tachyon and multi-field phantom tachyon known as the multi-field tachyon-quintom model. During evolution of the universe the equation of state parameter  $w$  in  $p=w\rho$  alters to  $w<-1$  from  $w>-1$  in this model. Inverse square potentials are used in the development of the autonomous system for doing the analysis in phase space where we found stable points that have power-law solutions. Analysis of spatially flat background universe of FLRW metric manifests the existence of a unique critical point which is compared with the tachyon dark energy model. We observed that neither multi-field tachyon nor the multi-field phantom tachyon showed sensitivity to the kinetic energy of initial conditions. It happens almost when the e-folding number is 8 and variation of multi-field tachyons by the order of magnitude four is still observed coinciding with the observations conducted in the recent past.

\end{document}